\documentclass[preprintnumbers,article,amsmath,amssymb,floatfix,10pt,prd,onecolumn,
superscriptaddress,nofootinbib]{revtex4}
\usepackage{bbm}
\usepackage{amsfonts}
\usepackage{mathrsfs}
\usepackage{latexsym}

\usepackage{epsfig}
\usepackage{epstopdf}
\usepackage{graphicx}
\usepackage{amssymb}
\usepackage{amsmath}
\usepackage{dcolumn}
\usepackage{bm}
\usepackage{color}
\usepackage{comment}
\usepackage{xcolor}

\begin{document}

\title{\bf Bardeen Stellar Structures with Karmarkar Condition}

\author{G. Mustafa}
\email{gmustafa3828@gmail.com}\affiliation{Department of Mathematics, Shanghai University,
Shanghai, 200444, Shanghai, People's Republic of China}
\author{M. Farasat Shamir}
\email{farasat.shamir@nu.edu.pk; farasat.shamir@gmail.com}\affiliation{National University of Computer and
Emerging Sciences,\\ Lahore Campus, Pakistan.}
\author{Mushtaq Ahmad}
\email{mushtaq.sial@nu.edu.pk}\affiliation{National University of Computer and
Emerging Sciences,\\ Chiniot-Faisalabad Campus, Pakistan.}

\begin{abstract}
This current study is focussed to discuss the existence of a new family of compact star solutions
by adopting the Karmarkar condition in the background of Bardeen black hole geometry. For this
purpose, we consider static spherically symmetric spacetime with anisotropic fluid distribution
in the presence of electric charge. We consider a specific model of $g_{rr}$ metric
function, to describe a new family of solutions which satisfies the Karmarkar condition.
Further, we investigate the interior solutions for two different models
of compact stars with observational mass and radii, i.e., $(M=1.77M_{\odot}, \;R_{b}=9.56km)$ and $(M=1.97M_{\odot}, \;R_{b}=10.3km)$. It is found that these solutions fulfill all the necessary
conditions for a charged star. Through graphical discussion, it is
noticed that our calculated solutions are physically arguable with a best degree of
accuracy for $n\in[1.8,7)$, where parameter $n$ is involved in the model under discussion.
However, it is perceived that the presented model violates all the physical conditions for $n\in\{2,4,6\}$.
Finally, it is concluded that the parameter $n$ has a strong impact on the obtained solutions in the context of Bardeen stellar structures.\\\\\\
\textbf{Keywords}: Bardeen Model; Stellar Structures; Karmarkar Condition.
\end{abstract}

\maketitle

\date{\today}

%%%%%%%%%%%%%%%%%%%%%%%%%%%%%%%%%%%%%%%%%%%%%%%%%%%%%%%%%%%%%%%%%%%%%%%%
%%%%%%%%%%%%%%%        Introduction        %%%%%%%%%%%%%%%%%%%%%%%%%%%%%
%%%%%%%%%%%%%%%%%%%%%%%%%%%%%%%%%%%%%%%%%%%%%%%%%%%%%%%%%%%%%%%%%%%%%%%%
\section{Introduction}

Compact structures-- neutron stars,  black holes, and white dwarfs are created when the normal stars "die," that is, when the most of their nuclear fuel has been spent. All these three types of compact object are quite different from normal stars in two ways of primarily concern. Firstly, they cannot hold themselves up against the gravitational collapse by producing thermal pressure as they do not consume nuclear fuel. On the other side, white dwarfs are reinforced by the strong pressure of degenerate electrons, whereas the neutron stars are strengthened to exist largely by the degenerate pressure due to neutrons. Black holes instead, are totally collapsed stellar remnants-that is, the compact objects that could not find any support sufficient to hold back the centripetal pull of gravity and consequently collapsed to the singularities. With the exclusion of some spontaneously radiating "small" black holes with masses $M$ less than $10^{15} g$ and radii lesser than a fermi, all three compact objects have remained fundamentally static  ever since the lifetime of the Universe. They signify the final phase of the stellar evolution. Secondly, the compact objects which are exclusively distinguished  from the standard stars is to possess the exceedingly small mass and size. Relative to these normal stars with  comparable masses, compact objects have much fewer radii and therefore very strong gravitational fields over the boundary. This dramatic phenomenon is caused due to enormous energy density range extends across the compact objects, and this requires a deep physical understanding for the detailed investigation of the complexion of the  matter involved and the nature of inter-particle cohesive forces over an enormous range of parametric space. All four vital interactions, that is the weak and the strong nuclear forces, gravitation, and electromagnetism, play their role in the formation of the compact objects. The large-scale surface potentials being encountered in compact objects are worth mentioning here, which suggest that the  general relativity (GR) is of great significance in building their structure. Even for white dwarfs, where Newtonian approach of gravitation is suitable to describe their equilibrium structure, GR proves to be worthy for a better understanding of their stability.\par

Compact objects like gravastars, neutron stars and quark stars have been
the focuss of attention of many researchers in the recent decades. Digging up the exact solutions of the
Einstein field equations had begun since the famous Schwarzschild vacuum solution for
spherically symmetric matter distribution \cite{1Schwarz1}.
A standard pursuit of such findings would be to quest for an interior solution that should match smoothly to the exterior region of the Schwarzschild solution.
Schwarzschild obtained this solution by assuming
that the interior matter structure of the
spherically symmetric distribution was described by the uniform
energy density \cite{2Schwarz2}. Some interesting observations of compact structures and for the purpose to develop the deep understanding of perceptions of the particle physics within the bounds of condensed matter compelled the investigators to hunt for some more realistic explanations of the field equations. As far as equation of state (EoS) is concerned, charge, pressure anisotropy, multilayered fluids and bulk viscosity have served the purpose of uncovering diverse exact solutions for the exploration of relativistic stars in the presence of static limit \cite{3Bowers}-\cite{4Bekenstei}. By the discovery of  Vaidya solution, it was deliberately  desired to conceptualize such model that could expose the gravitational breakdown of the radiating stellar structures \cite{8Vaidya}. Soon after the stars start emitting energy as radial heat flux form, the stellar surface pressure becomes proportional to the out-flowing heat flux which conflicts to the non-dissipative case when the surface pressure disappears \cite{Santos}. However,  a key role in dissipative gravitational collapse of stars is still eminent from the static solutions as they can characterize an early static structure or conclusive static configuration \cite{10Bonnor}-\cite{12Govende}.\par

It is worth noting here that by circumventing the perfect fluid condition and authorizing the charge within the interior through the anisotropic pressure of the stellar structures,  gives birth to noticeable stellar features. The inclusion of the electric charge provides the platform for  the adjustment of the Buchdahl limit, whereas the pressure anisotropy concludes into the arbitrary big surface redshifts \cite{13Bhar2}-\cite{16Andreasson}.
The general EoS with the expression $p=\rho\omega$  has arisen from the annotations of theoretical particle physics. A wide diversity of exact solutions of the field equations integrating with the eminent MIT bag model with the EoS expression as $p=\alpha\rho-\beta$, here $\beta$ stands for the bag model constant \cite{17Takisa}-\cite{19Govender1}. The solutions obtained so magnificently predicted the presumed masses and radius of the compact stars comprising the energy densities totaling to the value of $10^{14}g cm^{-3}$. With the ever since increasing dependency on the assumptions made on the dark energy and its useful implications in astrophysical models, scientists have now started to extend the range up to $-1<\alpha<-1/3$ in the context of $p=\alpha\rho$.  This consideration embraces so-called dark models of stellar structures \cite{20Bhar5}-\cite{22Rahaman4}. The exotic matter structures other than already noted in the literature includes Bose-Einstein condensates, Chaplygin gas, and the Hagedorn fluid \cite{23Rahaman5}-\cite{27Dadhich}.\par

In order to investigate physically stable models, we need to find an analytical approach of the Einstein field equations.
An important direction is to use the embedding class of a four dimensional manifold into a higher dimensional Euclidean space. Embedding class of curved spacetimes into spacetimes of higher dimensions is assumed to generate several new exact models in cosmology and relativistic astrophysics. The embedding class condition provides a differential equation in spherically static spacetimes connecting the two gravitational potentials, known as Karmarkar condition \cite{34Karmarkar3}. This condition has been proved to be an important and fruitful mechanism to find new solutions of relativistic astrophysical models.
Schlai \cite{E1} was the fisrt who addressed the embedding problem on geometrically significant spacetimes. Later on, Nash \cite{E2} gave the isometric embedding
theorem. Gupta and Gupta \cite{E3} obtained a family of non-static fluid spheres of class one with non-vanishing acceleration. In another paper, Gupta and Sharma \cite{E4} studied non-static perfect fluid solutions by considering a plane symmetric metric of embedding class one.
Maurya et al. \cite{M1}-\cite{M6} were among the pioneers to study different aspects of anisotropic compact stars using embedding class one approach.
Higher-dimensional gravitational theories have come up with the vital results in the epoch of cosmic restriction \cite{28Joshi4}-\cite{30Hansraj1}. A big surge for finding the stellar exact solutions have started in recent years, just to mention few, in Gauss-Bonnet gravity, Lovelock gravity and the braneworld gravity \cite{31Maharaj3}-\cite{33Banerjee2}. The relation between five-dimensional Kaluza-Klein geometries and the electromagnetism and has been widely investigated. Taking into consideration the four-dimensional spacetimes with higher degree geometrical models has remained a valuable tool in constructing both the astrophysical and the cosmological models.

The significance of Schwarzschild solution \cite{2Schwarz2} was the exploration of spacetime singularity from where the primary concept of a black hole originated. Whereas other non-trivial solutions anticipated the bounded compactness configuration parameter  that is $\mu{(R)}=\frac{2M(R)}{R}<\frac{8}{9}$ for the static, spherically symmetric hydrostatic equilibrium \cite{Buchdahl}.  Baade and Zwicky \cite{Baade} investigated the compact stars and commented that supernova might turn to the shape of some compact object with fewer density, and this was proven correct  after the discovery of pulsars which are believed to be  massively magnetized rotating neutrons \cite{Longair, Ghosh}. The work showing that nuclear density goes anisotropic at the interior of the compact stars, came from Ruderman \cite{Ruderman}. A big range of investigations has been initiated to explore the solutions of the field equations in different contexts \cite{Maurya}-\cite{Maharaj}. It is believed that pressure of the stellar fluid sphere splits up to the tangential and radial pressure in anisotropic geometries. Some further surveys unveil that the repulsive forces coming form the compact stars, are born due to anisotropic behavior. Kalam et al. \cite{Kalam} discussed the necessary conditions provided by Krori and Barua metric \cite{K&B} to uphold an effective and stable approach towards the modelling of relativistic compact objects. By structuring a consistent Tolman Oppenheimer Volkoff (TOV) equation, the numerical simulations may be considered to study the essentials of the compact stars, via the EoS parameter. Rahaman et al. \cite{Rahaman1, Rahaman2} investigated the EoS Chaplygin gas to uncover their physical possessions by extending the Krori and Barua models. Mak and Harko \cite{MaK} investigated few standard models for spherically symmetric stellar remnants and proposed exact solutions to determine the physical parameters such as  energy density, tangential and radial pressure terms with the deduction that inside these compact objects, the vital parameters would keep on finitely positive.

In this study, we are focussed to discuss the existence of a new family of charged compact star solutions.
The theoretical possibility of studying stellar models with an effect of
electric field has been done previously by many authors, for some references see (\cite{Bekenstein}-\cite{Takisa}).
Rosseland \cite{2600} discussed the possibility of a self gravitating star
treated as a ball of hot ionized gases containing a considerable amount of charge.
Thus large number of electrons in such a system as compared to positive ions, run to
escape from its surface due to their higher kinetic energy and the motion of electrons will
continue until the electric field remains induced inside the compact star. In this way, the equilibrium is attained
after some electrons escape and the net electric charge approaches to
about 100 Coulombs per solar mass. Thus the possibility of collapsing of a star to a point
singularity may be avoided by the effects of charge. So, in this paper we investigate the charged compact star solutions
by adopting the Karmarkar condition in the background of Bardeen black hole geometry. To best of our knowledge, this is the first such attempt.
For massive stellar objects, the radial pressure may not be equal to the tangential one. Different arguments have been given for the existence of anisotropy in stellar models such as by the presence of type $3$A superfluid \cite{8143} and different kinds of phase
transitions \cite{9143}. In fact, anisotropy is also important to understand the peculiar properties of matter in the core of stellar structure. Thus, we consider static spherically symmetric spacetime with anisotropic fluid distribution in the presence of electric charge. The layout of this paper is as follows: In section $\textbf{2}$, we present some important mathematics regarding basic Einstein-Maxwell field equations with Karmarkar condition. Section $\textbf{3}$ is devoted to provide matching conditions using Bardeen geometry. Section $\textbf{4}$ provides a detailed physical analysis of the work. Last section is based on the conclusive remarks.

\section{Basic Field Equations}

For the present study, we consider the static spherically symmetric
space-time as
\begin{equation}\label{1}
ds^{2}=e^{\lambda(r)}dr^2+r^{2}d\theta^{2}+r^2sin^{2}\theta d\phi^{2}-e^{\nu(r)}dt^{2}.
\end{equation}
The anisotropic source of energy-momentum tensor in the presence of charge is given as
\begin{equation}\label{2}
\mathcal{T}_{\chi\gamma}=(\rho+p_{t})\upsilon_{\chi}\upsilon_{\gamma}-p_{t}g_{\chi\gamma}+(p_{r}-p_{t})\xi_{\chi}\xi_{\gamma}
+\frac{1}{4 \pi} (-\mathcal{F}^{\zeta \chi} \mathcal{F}_{\eta \chi} + \frac{1}{4} \delta^{\zeta}_{\eta} \mathcal{F}^{\chi \psi} \mathcal{F}_{\chi \psi} ),
\end{equation}
where $p_{r}$ and  $p_{t}$ are the radial, and tangential components of pressure source, $\rho$ is an energy density source and $\mathcal{F}^{\zeta \chi}$ is the usual Maxwell's stress tensor.
The four velocity vector is denoted by $\upsilon_{\chi}$ and the radial four vector by $\xi_{\alpha}$, satisfying the following condition
\begin{equation*}
\upsilon^{\alpha}=e^{\frac{-\nu}{2}}\delta^{\alpha}_{0},~~~\upsilon^{\alpha}\upsilon_{\alpha}=1,~~~\xi^{\alpha}=e^{\frac{-\lambda}{2}}\delta^{\alpha}_{1},~~~\xi^{\alpha}\xi_{\alpha}=-1.
\end{equation*}
Einstein-Maxwell's field equations (assuming the gravitational units) are given by
\begin{eqnarray}\nonumber
R_{\mu\nu}-\frac{1}{2} R g_{\mu\nu} &=& -8\pi T_{\mu\nu},\\\nonumber
\mathcal{F}^{\eta\zeta}_{;\zeta}&=&-4 \pi j^\eta,\\\label{3}
\mathcal{F}_{[\beta\zeta;\eta]}&=&0,
\end{eqnarray}
where $j^\eta$ is the electromagnetic four current vector. Maxwell's stress tensor and electromagnetic four current vector are defined as
\begin{eqnarray}\nonumber
\mathcal{F}^{\eta \zeta}&=&A_{\zeta,\eta}-A_{\eta, \zeta},\\\label{4}
j^\eta&=&\sigma \nu^{\eta},
\end{eqnarray}
respectively. Here $A$ being the magnetic four potential and $\sigma$ symbolizes as charge density.
The only non-vanishing component in the static spherically symmetric system is $J^{0}$. The Einstein-Maxwell tensor contains the only non-zero component $\mathcal{F}^{01}$ defined as
\begin{eqnarray}\label{5}
\mathcal{F}^{01}=-\mathcal{F}^{10}=\frac{q}{r^2} e^{-(\frac{\nu+\lambda}{2})},
\end{eqnarray}
where the term $q$ represents the charge inside the spherical stellar system and is given by
\begin{eqnarray}\label{6}
q=4\pi\int_{0}^{r} 4 \sigma \rho^2 e^{(\frac{\lambda}{2})} d\rho.
\end{eqnarray}
The term electric field intensity $E$ can be expressed as
\begin{eqnarray}
~~~~~~~~~~E^2=-\mathcal{F}^{01}\mathcal{F}_{10}=\frac{q^2}{ r^4}.~~~~~~~~~~\quad\quad~~~~~~\label{7}
\end{eqnarray}
Now for the spacetime (\ref{1}) along with the matter distribution (\ref{2}), Einstein-Maxwell's field equations (\ref{4}) turn out to be
\begin{eqnarray}\label{8}
8\pi\rho+E^2&=&\frac{1}{ e^{\lambda}r^2}\big(\lambda'r+e^{\lambda}-1\big),
\\\label{9}
8\pi p_r-E^2&=&\frac{1}{ e^{\lambda}r^2}\big(\nu'r-e^{\lambda}+1\bigg),~\\\label{10}
8\pi p_t+E^2&=&\frac{1}{ e^{\lambda}}\big(\frac{{\nu'}^{2}}{4}+\frac{\nu''}{2}-\frac{\nu'\lambda'}{4}+\frac{\nu'}{2r}-\frac{\lambda'}{2r}\big),\\\label{11}
 \sigma&=& \frac{e^{-\lambda/2}}{4\pi r^2}(r^2E)'.
\end{eqnarray}
Now, we shall explore the well-known condition, i.e., Karmarker condition \cite{34Karmarkar3} which is one of the most important aspect of the present study. The formation of Karmarkar condition depends upon the embedded Riemannian-space of class-I. Eisenhart \cite{41} calculated a necessary and sufficient condition based on a symmetric tensor of second order, i.e., $\chi_{\nu\eta}$ and the Riemann curvature tensor $R_{\nu \lambda\eta\gamma}$, which is discussed as
\begin{eqnarray*}
\Sigma(\chi_{\nu\eta}\chi_{\lambda\gamma}-\chi_{\nu\gamma}\chi_{\lambda\eta})&=&R_{\nu \lambda\eta\gamma},\\
\chi_{\nu \lambda;\eta}-\chi_{\nu\eta;\lambda}&=&0,
\end{eqnarray*}
where $\verb";"$ mentions the covariant derivative and $\Sigma=\pm 1$ shows that the manifold is a time-like or a space-like.
The Riemann tensor components for embedded class-1, are calculated as
\begin{eqnarray*}
R_{1414}&=&\frac{e^{\nu (r)}(2\nu ''(r)+\nu '(r)^2-\nu '(r)\lambda'(r))}{4},\;\;\;\;\;R_{2323}=\frac{r^{2}sin^{2}\theta (e^{\lambda (r)}-1)}{e^{\lambda (r)}},\nonumber\\
R_{1212}&=&\frac{r \lambda '(r)}{2},\;\;\;\;\;\;\;\;\;\;\;\;\;\;\;\;\;\;\;\;\;\;\;\;\;\;\;\;\;\;\;\;\;\;\;\;\;\;\;\;\;\;\;\;\;\;\;\;\;R_{3434}=\frac{rsin^{2}\theta \lambda '(r) e^{\nu (r)-\lambda (r)}}{2},\nonumber\\
R_{1334}&=&R_{1224}sin^{2}\theta,\;\;\;\;\;\;\;\;\;\;\;\;\;\;\;\;\;\;\;\;\;\;\;\;\;\;\;\;\;\;\;\;\;\;\;\;\;\;\;\;\;\;R_{1224}= 0.\nonumber
\end{eqnarray*}
Now, by plugging these non-zero Riemann tensor components for Eq. (\ref{1}) under the non-zero components of second order of symmetric tensor $\chi_{a\eta}$, we get an equation as:
\begin{equation}\label{16}
R_{1414}R_{2323}=R_{1224}R_{1334}+ R_{1212}R_{3434},
\end{equation}
where, $R_{2323}\neq0$, then it represents the spacetime of
emending class one.
The Karmarkar condition leads us to the following differential equation
\begin{equation}\label{13}
\lambda^{'}(r)\nu^{'}(r)+\nu^{'^{2}}(r)-2(\nu^{''}(r)+\nu^{'^{2}}(r))=\frac{\nu^{'}(r)\nu^{'}(r)}{1-e^{\lambda}},
\end{equation}
with $e^{\nu(r)}\neq1$. On solving the Eq. (\ref{13}), we get the following relation
\begin{equation}\label{14}
e^{\nu(r)}=\left(B\int(\sqrt{e^{\lambda(r)}-1})dr+A\right)^{2},
\end{equation}
where, $B$ and $A$ are the arbitrary parameters.
The anisotropy profile, i.e., $\triangle$ can be measured from the difference of tangential and radial pressure, i.e., $p_{t}-p_{r}$ given as
\begin{equation}\label{new}
\triangle=8\pi( p_{t}-p_{r})=\frac{\nu '(r)}{4 e^{\lambda(r)}}\left(\frac{2}{r}-\frac{\lambda'(r)}{e^{\lambda(r)}-1}\right) \left(\frac{e^{\nu (r)} \nu '(r)}{2 B^2 r}-1\right)-2E^{2}.
\end{equation}
Let us ansatz the other metric component $g_{rr}$, which is defined as:
\begin{equation}\label{15}
e^{\lambda(r)}=\frac{c r^2 \left(X r^2+1\right)^n}{\left(Y r^2+1\right)^2}+1,
\end{equation}
where $c$, $X$, $Y$ and $n$ are the free parameters. It is worthwhile to mention here that all the analysis in this work depends on the metric functions, i.e., $g_{tt}=e^{\nu(r)}$ and $g_{rr}=e^{\lambda(r)}$. Lake \cite{1000} has argued that for any physically valid solution, the metric potentials must be a positive, monotonically increasing functions of the radial coordinate and regular throughout the stellar model.
%More importantly, Eq. (\ref{15}) satisfies the Karmarkar condition and also provides the metric function free from any geometrical singularity.
The interesting feature of class-one condition is that the both metric functions are dependent on each other. In this situation, we will get only two
types of perfect-fluid solutions either Schwarzschild solution \cite{2Schwarz2} or Kholar-Chao solution \cite{KC}, if anisotropy is zero.
However, we can introduce the anisotropy or charge in the system through the class one condition only if the considering solution is different from Kohler-Chao or Schwachild solution \cite{M1,M2}.
Manipulating Eqs. (\ref{14}) and (\ref{15}), we get required metric coefficient $g_{tt}$ as:
\begin{equation}\label{16}
e^{\nu(r)}=\left(\frac{B \left(Y r^2+1\right) \left(\frac{X Y r^2+Y}{X Y r^2+X}\right)^{-\frac{n}{2}} \sqrt{\frac{c r^2 \left(X r^2+1\right)^n}{\left(Y r^2+1\right)^2}} F(r)}{Y n r}+A\right)^2,
\end{equation}
where
\begin{equation*}
F(r)=\, _2F_1\left(-\frac{n}{2},-\frac{n}{2};1-\frac{n}{2};\frac{X-Y}{X Y r^2+X}\right).
\end{equation*}
Here we introduce an electric field of the form $E^2 =KQr$. Further, by plugging the values of $e^{\nu(r)}$ and $e^{\lambda(r)}$, from Eqs. (\ref{15}) and (\ref{16}) in field Eqs. (\ref{8})-(\ref{11}), it follows
\begin{eqnarray}
8\pi\rho&=&-\frac{c F_0(r) \left(X r^2+1\right)^{n-1}}{\left(c r^2 \left(X r^2+1\right)^n+\left(Y r^2+1\right)^2\right)^2}+K Q r,\label{17}\\
8\pi p_r&=&\frac{r \left(B c r F(r) F_4(r) \left(X r^2+1\right)^n+Y n \left(Y r^2+1\right) F_3(r) F_2(r)^{n/2}\right)}{\left(Y r^2+1\right) \sqrt{F_1(r)} F_5(r) \left(c r^2 \left(X r^2+1\right)^n+\left(Y r^2+1\right)^2\right)},\label{18}\\
8\pi p_t&=&-\frac{r \left(B c r F(r) F_7(r) \left(X r^2+1\right)^n+Y n \left(Y r^2+1\right) F_9(r) F_2(r)^{n/2}\right)}{\left(X r^2+1\right) \left(Y r^2+1\right) \sqrt{F_1(r)} F_6(r){}^2 F_5(r)},\label{19}\\
8 \pi\sigma&=& \frac{2 \left(\frac{K Q r^2}{2 \sqrt{K Q r}}+2 r \sqrt{K Q r}\right)}{r^2 \sqrt{F_1(r)+1}},\label{20}
\end{eqnarray}
where $F_i(r)$, $\{i=0,...,9\}$ are given in the Appendix (\textbf{I}).

\section{Matching Conditions}

In this section, we propose Bardeen model to describe as an exterior
line element given by \cite{Bardeen}
\begin{equation}\label{21}
ds^{2}={f(r)}^{-1}dr^2+r^{2}d\theta^{2}+r^2sin^{2}\theta d\phi^{2}-f(r)dt^{2},
\end{equation}
where $f(r)=1 -\frac{2mr^2}{({q^2}+{r^2})^{\frac{3}{2}}}$. It has been proved that the Bardeen black hole can be interpreted
as a gravitationally collapsed magnetic monopole arising from some specific case
of non-linear electrodynamics \cite{Garcia}. Moreover, Bardeen black holes can be obtained as exact solutions of some
appropriate non-linear electrodynamics coupled to gravity and the non-zero Einstein tensor in the Bardeen model can be associated with the stress-energy tensor
of a nonlinear electromagnetic Lagrangian \cite{Moreno}. Further, the existence of Bardeen black solutions does not contradict the singularity theorems \cite{Hawking}. The discussion of Bardeen black hole has attracted much attention in different contexts \cite{Fernando1}-\cite{Ulhoa}. It is worthwhile to notice that the spacetime asymptotically behaves as
\begin{equation}\label{22}
f(r) = 1-\frac{2M}{r}+\frac{3Mq^2}{r^3} + O(\frac{1}{r^5}).
\end{equation}
One can notice that the term $1/r$ involved in Eq. (\ref{22}) suggests that the parameter $M$ is associated with the mass of the stellar configuration. This fact can also be proved from the explicit derivation of the ADM mass formalism. However, the next term involves $1/r^3$ which makes the things more interesting as this case this does not associate the parameter $q$ with some kind of "Coulomb" charge as it happens in the Reissner–
Nordstrom solution \cite{Nordstrom}. Thus in present study we consider $f(r)\approx 1-\frac{2M}{r}+\frac{3Mq^2}{r^3}$. Motivated from the above mentioned intersecting discussion, it would be an interesting task to study compact stars with Bardeen black hole geometry. In particular, the term involved $\frac{3Mq^2}{r^3}$ in Bardeen model corresponding to the the term $\frac{q^2}{r^2}$ in the usual Reissner Nordstrom model may provide some fascinating results.

Now, at the boundary $r={R_b}$ by junction condition under the continuity of the metric components $g_{tt}$ and $g_{rr}$, which are defined as,
\begin{equation}\label{23}
g_{tt}^-=g_{tt}^+,\;\;\;\;\;\;\;\;\;g_{rr}^-=g_{rr}^+,
\end{equation}
where $-$ and $+$, correspond to interior and exterior solutions.
\begin{eqnarray}
1-\frac{2M}{{R_b}}+\frac{3Mq^2}{{R_b}^3}&=&\frac{c r^2 \left(X r^2+1\right)^n}{\left(Y r^2+1\right)^2}+1,\label{24}\\
\bigg(1-\frac{2M}{{R_b}}+\frac{3Mq^2}{{R_b}^3}\bigg)^{-1}&=&\left(A+\frac{B \left(Y r^2+1\right) F(r) \sqrt{F_1(r)} F_2(r){}^{-\frac{n}{2}}}{Y n r}\right)^2.\label{25}
\end{eqnarray}
Being physically acceptable stellar model the radial component of pressure must be zero at the
boundary, which is mentioned as:
\begin{equation}\label{26}
p_r(r = R_b)= 0.
\end{equation}
Manipulating Eqs. (\ref{24}-\ref{26}), we get the expressions for the parameter $c,\;A$, and $B$ as
\begin{eqnarray}
c&=&-\frac{M \left(Y R_{b}^2+1\right)^2 \left(X R_{b}^2+1\right)^{-n} \left(3 K Q R_{b}^3-2\right)}{R_{b}^2 \left(M \left(3 K Q R_{b}^3-2\right)+R_{b}\right)},\label{27}\\
A&=&-\frac{B H(R_{b}) H_2(R_{b}){}^{-\frac{n}{2}} \sqrt{c \left(X R_{b}^2+1\right)^n}}{Y n}-\frac{\sqrt{M \left(3 K Q R_{b}^3-2\right)+R_{b}}}{\sqrt{R_{b}}},\label{28}\\
B&=&-\frac{Y n \left(Y R_{b}^2+1\right) \sqrt{H_1(R_{b})} H_3(R_{b}) H_2(R_{b})^{n/2} \sqrt{M \left(3 K Q R_{b}^3-2\right)+R_{b}}}{\sqrt{R_{b}} \left(H_3(R_{b}) H(R_{b}) H_1(R_{b})-2 Y c n R_{b} \left(Y R_{b}^2+1\right) \left(X R_{b}^2+1\right)^n H_2(R_{b})^{n/2}\right)},,\label{29}
\end{eqnarray}
where
\begin{eqnarray*}
H(R_{b})&&=\, _2F_1\left(-\frac{n}{2},-\frac{n}{2};1-\frac{n}{2};\frac{X-Y}{X Y R_{b}^2+X}\right),\\
H_1(R_{b})&&=\left(Y R_{b}^2+1\right) \sqrt{F_1(R_{b})} \sqrt{c \left(X R_{b}^2+1\right)^n}-c R_{b} \left(X R_{b}^2+1\right)^n,\\
H_2(R_{b})&&=\frac{X Y R_{b}^2+Y}{X Y R_{b}^2+X},\\
H_3(R_{b})&&=c \left(X R_{b}^2+1\right)^n \left(K Q R_{b}^3-1\right)+K Q R_{b} \left(Y R_{b}^2+1\right)^2.\\
\end{eqnarray*}
Here $K,\;Q$ and $n$ are free parameters, while $M$ and $R_{b}$, represent the mass and radius of the object respectively. In this study, we use two different models with different mass and radius, i.e., $(M=1.77, \;R_{b}=9.56)$ and $(M=1.97, \;R_{b}=10.3)$ with $X=0.001$, $Y=0.0015$, $K=0.0001$, and $Q=0.005$.
\begin{center}
\begin{table}[h]
\caption{\label{tab1}{Approximated values of $c,\; A$ and $B$ under $(M=1.77M_{\odot}, \;R_{b}=9.56km)$ with $X=0.001(km^{-2})$, $Y=0.0015(km^{-2})$, $K=0.0001$, and $Q=0.005$ for different suitable values of $n$.}}
\vspace{0.3cm}\begin{tabular}{|c|c|c|c|c|c|c|}
\hline
$n$ & $c$ & $A (km^{-2})$ &  $B(km^{-1})$\\
\hline
 1.80    \;\;\;\;\; \;\;   &0.00710006698968945    &-2.768582339414147    &-0.03177821125661471\\
 2.20     \;\;\;\;\; \;\;  &0.00685598541857070    &2.9151465072316550    &-0.03177821125661471\\
 2.60    \;\;\;\;\; \;\;   &0.00662029472791072    &0.7833263663224741    &-0.03177821125661472\\
 3.00    \;\;\;\;\; \;\;   &0.00639270646137687    &0.2371149793117672    &-0.03177821125661472\\
 3.40     \;\;\;\;\; \;\;  &0.00617294207900410    &-0.130426856790480    &-0.03177821125661472\\
 3.80   \;\;\;\;\; \;\;    &0.00596073261629601    &-0.888623763975047    &-0.03177821125661473\\
 4.20    \;\;\;\;\; \;\;   &0.00575581835504860    &0.7988963439348606    &-0.03177821125661473\\
 4.60    \;\;\;\;\; \;\;   &0.00555794850548039    &0.1078102544144219    &-0.03177821125661474\\
 5.00   \;\;\;\;\; \;\;    &0.00536688089930361    &-0.092601630362715    &-0.03177821125661474\\
 5.40   \;\;\;\;\; \;\;    &0.00518238169333675    &-0.233468269417885    &-0.03177821125661474\\
 5.80  \;\;\;\;\; \;\;    &0.00500422508330988    &-0.489702255511396    &-0.03177821125661475\\
 6.20  \;\;\;\;\; \;\;     &0.00483219302750817    &0.0042170775091966    &-0.03177821125661475\\
 6.60   \;\;\;\;\; \;\;    &0.00466607497991566    &-0.225733288060495    &-0.03177821125661475\\
 7.00 \;\;\;\;\; \;\;      &0.00450566763253295    &-0.303664392633107    &-0.03177821125661475\\
  \hline
\end{tabular}
\end{table}
\end{center}
\begin{center}
\begin{table}[h]
\caption{\label{tab1}{Approximated values of $c,\; A$ and $B$ under $(M=1.97M_{\odot}, \;R_{b}=10.3km)$ with $X=0.001(km^{-2})$, $Y=0.0015(km^{-2})$, $K=0.0001$, and $Q=0.005$ for different suitable values of $n$.}}
\vspace{0.3cm}\begin{tabular}{|c|c|c|c|c|c|c|}
\hline
$n$ & $c$ & $A (km^{-2})$ &  $B(km^{-1})$ \\
\hline
 1.80    \;\;\;\;\; \;\;   &0.00653481578396381    &-2.555859952141130    &-0.02996836039761251\\
 2.20     \;\;\;\;\; \;\;  &0.00627649465002195    &2.5773929172584316    &-0.02996836039761251\\
 2.60    \;\;\;\;\; \;\;   &0.00602838494520787    &0.6499999885619243    &-0.02996836039761252\\
 3.00    \;\;\;\;\; \;\;   &0.00579008301193752    &0.1559624090895727    &-0.02996836039761253\\
 3.40     \;\;\;\;\; \;\;  &0.00556120114920291    &-0.175518459683963    &-0.02996836039761254\\
 3.80   \;\;\;\;\; \;\;    &0.00534136698180893    &-0.853976150622281    &-0.02996836039761254\\
 4.20    \;\;\;\;\; \;\;   &0.00513022285454423    &0.6486676794313937    &-0.02996836039761254\\
 4.60    \;\;\;\;\; \;\;   &0.00492742525030075    &0.0312105081065342    &-0.02996836039761254\\
 5.00   \;\;\;\;\; \;\;    &0.00473264423119460    &-0.148429723097281    &-0.02996836039761253\\
 5.40   \;\;\;\;\; \;\;    &0.00454556290177967    &-0.274525095852821    &-0.02996836039761253\\
 5.80  \;\;\;\;\; \;\;     &0.00436587689348037    &-0.501572297658238    &-0.02996836039761253\\
 6.20  \;\;\;\;\; \;\;     &0.00419329386940462    &-0.068404386785546    &-0.02996836039761254\\
 6.60   \;\;\;\;\; \;\;    &0.00402753304873172    &-0.271769606034447    &-0.02996836039761254\\
 7.00 \;\;\;\;\; \;\;      &0.00386832474990105    &-0.341217822548745    &-0.02996836039761254\\
\hline
\end{tabular}
\end{table}
\end{center}
\section{Physical Analysis}
In this section, we analyze the physical parameters like, metric components, energy density, pressure components and their gradients with respect to radial coordinate.
\subsubsection{Metric potential and energy density}
Both the gravitational metric functions, i.e., $g_{tt}\;\&\;g_{rr}$ have key role in the compact stars study. Here in this study, we discuss an important model, i.e., $e^{\lambda(r)}=\frac{c r^2 \left(X r^2+1\right)^n}{\left(Y r^2+1\right)^2}+1$, we calculate the second metric component by employing the Karmarkar condition. It is calculated $e^{\lambda(r =0)}=1$ and $e^{\nu(r =0)}\neq0$, which justifies that the presence of Karmarkar relation is physically viable for this model. The convincing deviation of both the metric components can be reviewed from Fig. (\textbf{1}) for both the models under the different values of parameter $n$, which can be recognized from the caption of Fig. (\textbf{1}). We plot the energy density, its descriptive innovation can be identified from Fig. (\textbf{2}) for both the models by left and right plots. The decreasing trend of energy density shows the superiority of this model. At the center of two astral objects charged energy density is realized maximum, while on the boundary, it is examined minimum, which shows the acceptability of the model with highly accuracy.
\begin{figure}
\centering \epsfig{file=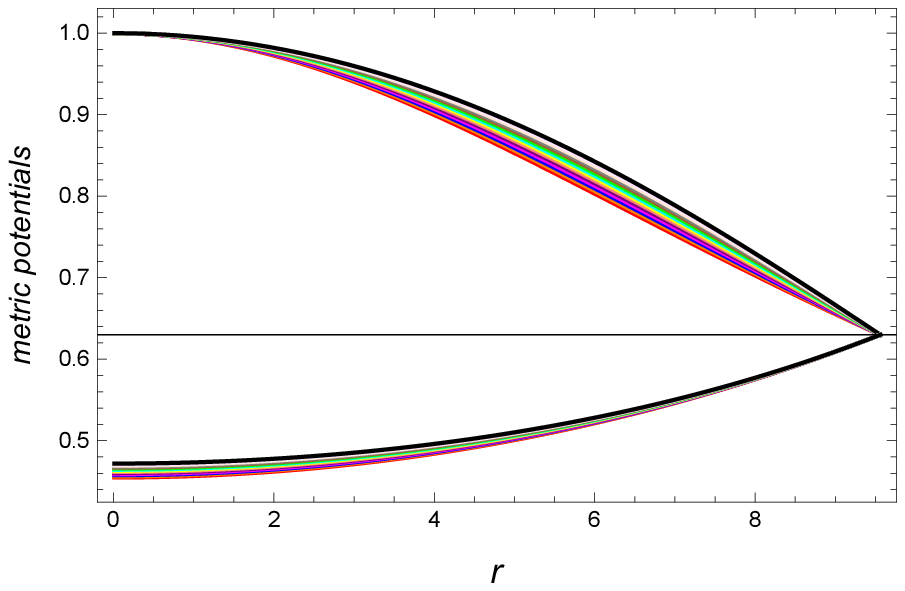, width=.48\linewidth,
height=2.1in}\epsfig{file=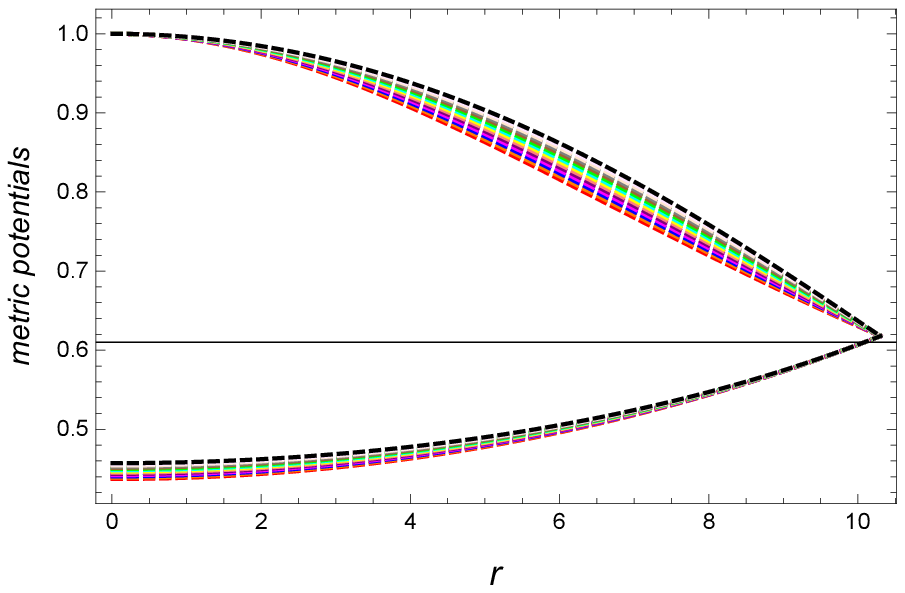, width=.48\linewidth,
height=2.1in}\caption{\label{Fig.1} Displays the evolution of metric functions for two different models under $(M=1.77, \;R_{b}=9.56)$ and $(M=1.97, \;R_{b}=10.3)$ with $n=1.80(\textcolor{red}{\bigstar})$, $n=2.20(\textcolor{orange}{\bigstar})$, $n=2.60(\textcolor{blue}{\bigstar})$, $n=3.00(\textcolor{magenta}{\bigstar})$, $n=3.40(\textcolor{purple}{\bigstar})$, $n=3.80(\textcolor{pink}{\bigstar})$, $n=4.20(\textcolor{yellow}{\bigstar})$, $n=6.60(\textcolor{cyan}{\bigstar})$, $n=5.00(\textcolor{green}{\bigstar})$, $n=5.40(\textcolor{brown}{\bigstar})$, $n=5.80(\textcolor{gray}{\bigstar})$,
$n=6.20(\textcolor{orange!50}{\bigstar})$, $n=6.60(\textcolor{magenta!50}{\bigstar})$, and $n=7.00(\textcolor{black}{\bigstar})$.}
\end{figure}
\begin{figure}
\centering \epsfig{file=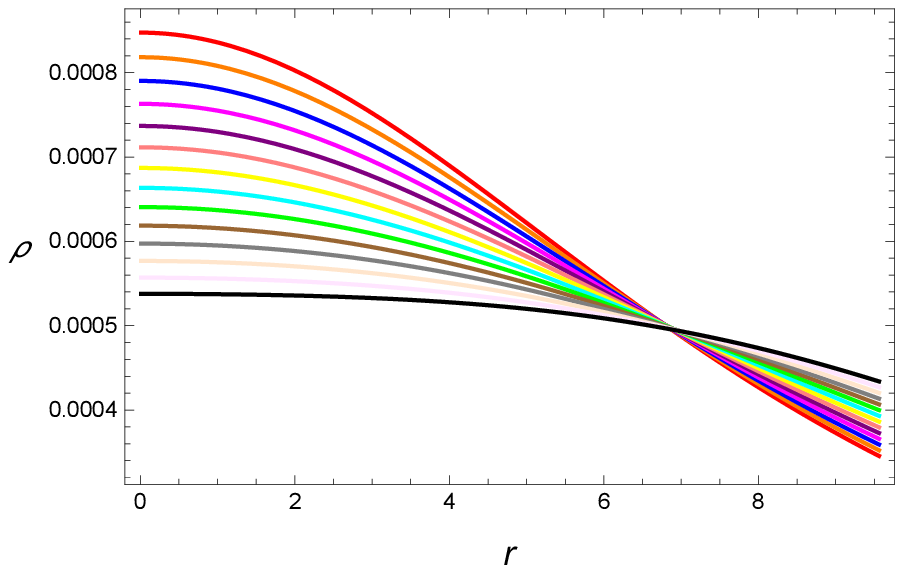, width=.48\linewidth,
height=2.1in}\epsfig{file=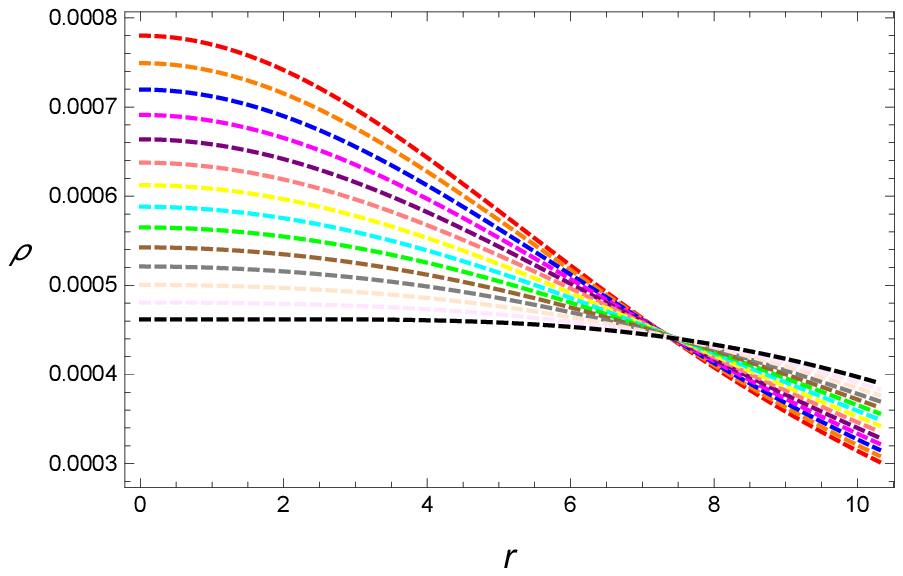, width=.48\linewidth,
height=2.1in}\caption{\label{Fig.2} Illustrates the evolution of energy density for two different models under $(M=1.77, \;R_{b}=9.56)$ and $(M=1.97, \;R_{b}=10.3)$ with $n=1.80(\textcolor{red}{\bigstar})$, $n=2.20(\textcolor{orange}{\bigstar})$, $n=2.60(\textcolor{blue}{\bigstar})$, $n=3.00(\textcolor{magenta}{\bigstar})$, $n=3.40(\textcolor{purple}{\bigstar})$, $n=3.80(\textcolor{pink}{\bigstar})$, $n=4.20(\textcolor{yellow}{\bigstar})$, $n=6.60(\textcolor{cyan}{\bigstar})$, $n=5.00(\textcolor{green}{\bigstar})$, $n=5.40(\textcolor{brown}{\bigstar})$, $n=5.80(\textcolor{gray}{\bigstar})$,
$n=6.20(\textcolor{orange!50}{\bigstar})$, $n=6.60(\textcolor{magenta!50}{\bigstar})$, and $n=7.00(\textcolor{black}{\bigstar})$.}
\end{figure}
\subsubsection{Pressure components}
Here, we discuss the radial and tangential sources of pressure profile. The non-negative radial and transverse pressure sources inside the stars is a compulsory requirement of compact stars study. In this study, the radial pressure presence is reported monotonically decreasing and it becomes zero at the boundary. Its graphic conduct can be examined from the Fig. (\textbf{3}) for two different stellar configuration. The tangential pressure is regarded as positive throughout the configuration. The Fig. (\textbf{4}) expresses the graphical analysis of transverse source of pressure.
\begin{figure}
\centering \epsfig{file=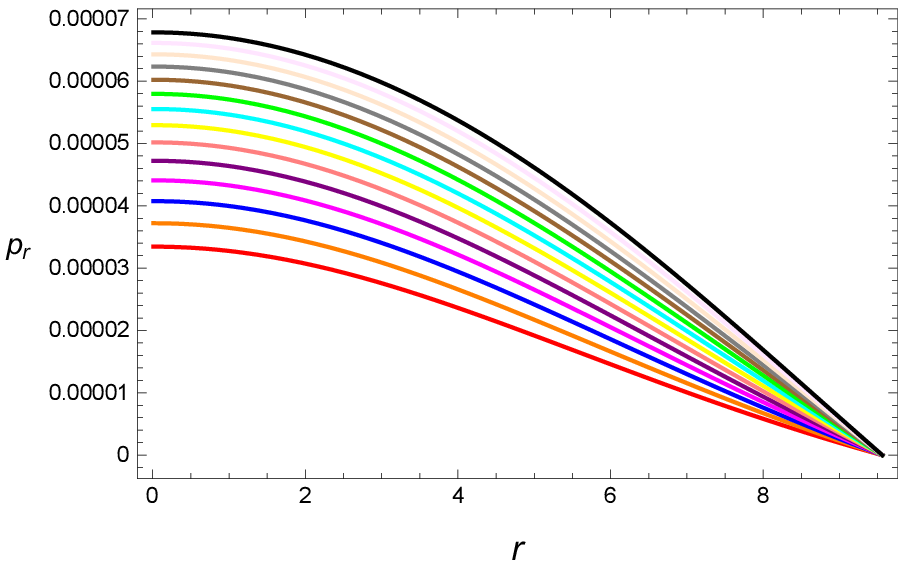, width=.48\linewidth,
height=2.1in}\epsfig{file=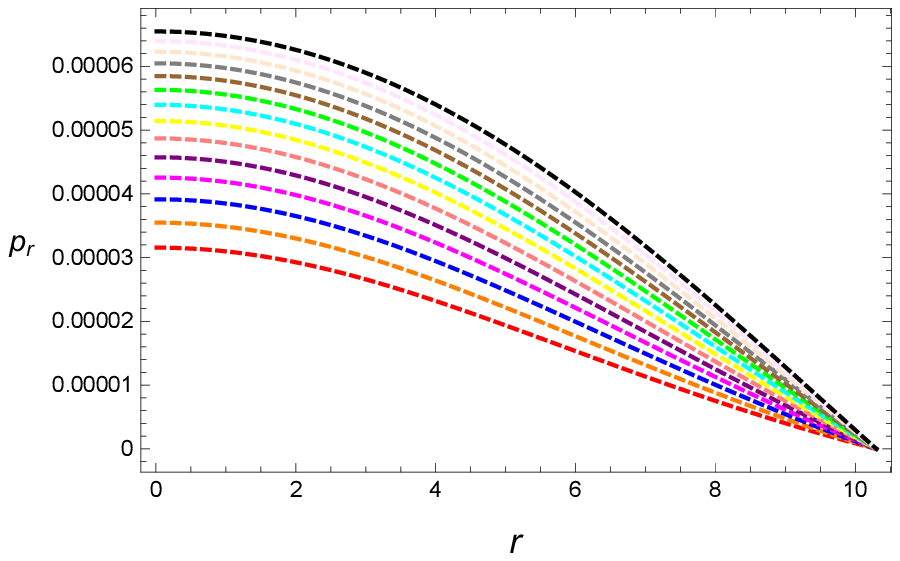, width=.48\linewidth,
height=2.1in}
\caption{\label{Fig.3} Shows the behavior of radial pressure for two different models under $(M=1.77, \;R_{b}=9.56)$ and $(M=1.97, \;R_{b}=10.3)$ with $n=1.80(\textcolor{red}{\bigstar})$, $n=2.20(\textcolor{orange}{\bigstar})$, $n=2.60(\textcolor{blue}{\bigstar})$, $n=3.00(\textcolor{magenta}{\bigstar})$, $n=3.40(\textcolor{purple}{\bigstar})$, $n=3.80(\textcolor{pink}{\bigstar})$, $n=4.20(\textcolor{yellow}{\bigstar})$, $n=6.60(\textcolor{cyan}{\bigstar})$, $n=5.00(\textcolor{green}{\bigstar})$, $n=5.40(\textcolor{brown}{\bigstar})$, $n=5.80(\textcolor{gray}{\bigstar})$,
$n=6.20(\textcolor{orange!50}{\bigstar})$, $n=6.60(\textcolor{magenta!50}{\bigstar})$, and $n=7.00(\textcolor{black}{\bigstar})$.}
\end{figure}
\begin{figure}
\centering \epsfig{file=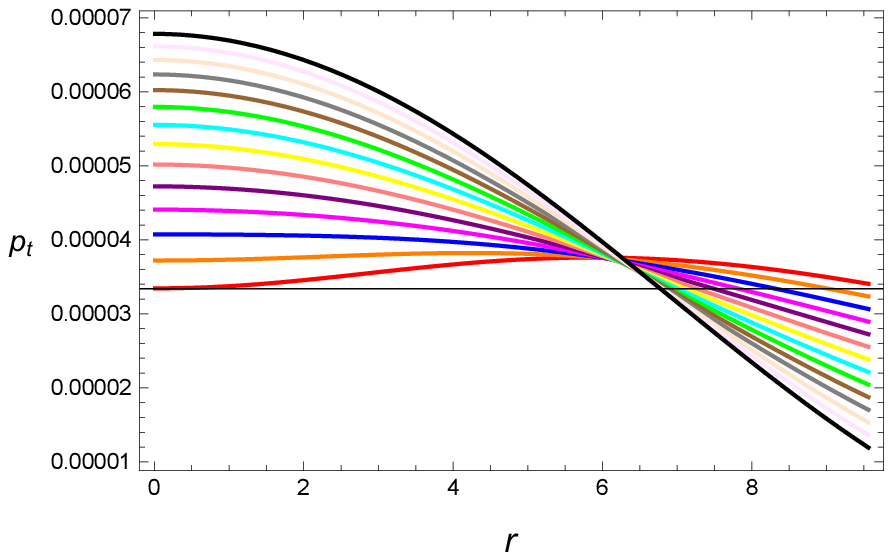, width=.48\linewidth,
height=2.1in}\epsfig{file=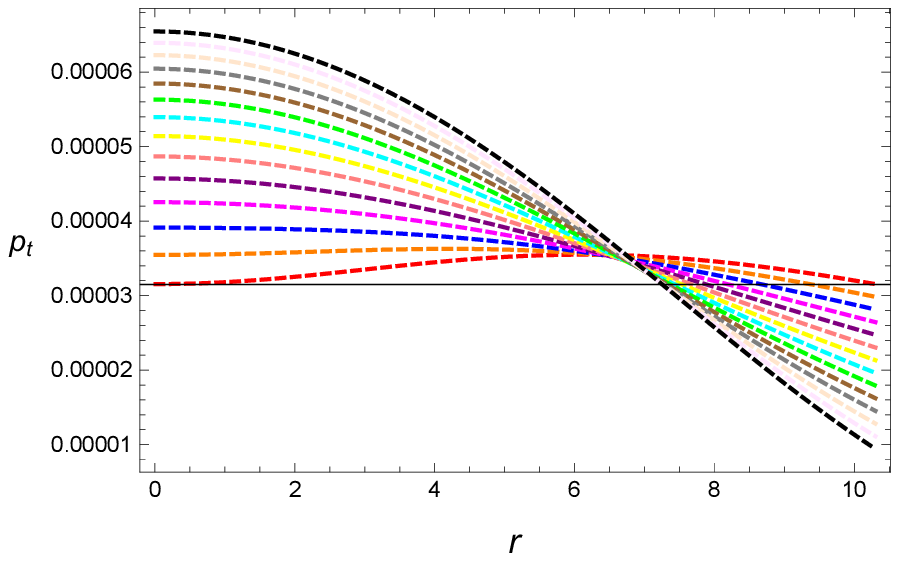, width=.48\linewidth,
height=2.1in}\caption{\label{Fig.4} Illustrates the development of tangential pressure for two different models under $(M=1.77, \;R_{b}=9.56)$ and $(M=1.97, \;R_{b}=10.3)$ with $n=1.80(\textcolor{red}{\bigstar})$, $n=2.20(\textcolor{orange}{\bigstar})$, $n=2.60(\textcolor{blue}{\bigstar})$, $n=3.00(\textcolor{magenta}{\bigstar})$, $n=3.40(\textcolor{purple}{\bigstar})$, $n=3.80(\textcolor{pink}{\bigstar})$, $n=4.20(\textcolor{yellow}{\bigstar})$, $n=6.60(\textcolor{cyan}{\bigstar})$, $n=5.00(\textcolor{green}{\bigstar})$, $n=5.40(\textcolor{brown}{\bigstar})$, $n=5.80(\textcolor{gray}{\bigstar})$,
$n=6.20(\textcolor{orange!50}{\bigstar})$, $n=6.60(\textcolor{magenta!50}{\bigstar})$, and $n=7.00(\textcolor{black}{\bigstar})$.}
\end{figure}
\subsubsection{Non-singular nature of the model}
There is one crucial point of this study that all the necessary physical quantities like, energy density, radial and tangential sources of pressure profile are distinguished non-singular.
\begin{eqnarray}
\rho_{c}&&=\frac{3c}{8\pi},\label{30}\\
p_{rc}&&=\frac{\sqrt{c} \left(Y n \left(\frac{Y}{X}\right)^{n/2} \left(2 B-A \sqrt{c}\right)-B c \, _2F_1\left(-\frac{n}{2},-\frac{n}{2};1-\frac{n}{2};1-\frac{Y}{X}\right)\right)}{8 \pi  \left(A Y n \left(\frac{Y}{X}\right)^{n/2}+B \sqrt{c} \, _2F_1\left(-\frac{n}{2},-\frac{n}{2};1-\frac{n}{2};1-\frac{Y}{X}\right)\right)},\label{31}\\
p_{tc}&&=\frac{\sqrt{c} \left(Y n \left(\frac{Y}{X}\right)^{n/2} \left(2 B-A \sqrt{c}\right)-B c \, _2F_1\left(-\frac{n}{2},-\frac{n}{2};1-\frac{n}{2};1-\frac{Y}{X}\right)\right)}{8 \pi  \left(A Y n \left(\frac{Y}{X}\right)^{n/2}+B \sqrt{c} \, _2F_1\left(-\frac{n}{2},-\frac{n}{2};1-\frac{n}{2};1-\frac{Y}{X}\right)\right)}.\label{32}
\end{eqnarray}
Further, the Zeldovich's condition, i.e., $\frac{p_{rc}}{\rho_{c}}$ or $\frac{p_{tc}}{\rho_{c}}$ is reported less than 1. The Zeldovich's condition is evaluated for this current study with electric charge as:
\begin{equation}\label{33}
\frac{Y n \left(\frac{Y}{X}\right)^{n/2} \left(2 B-A \sqrt{c}\right)-B c \, _2F_1\left(-\frac{n}{2},-\frac{n}{2};1-\frac{n}{2};1-\frac{Y}{X}\right)}{3 A Y \sqrt{c} n \left(\frac{Y}{X}\right)^{n/2}+3 B c \, _2F_1\left(-\frac{n}{2},-\frac{n}{2};1-\frac{n}{2};1-\frac{Y}{X}\right)}<1.
\end{equation}

\subsubsection{Charge density and electric field}

Here, we express the charge density and electric field. The charge density in both cases is recognized positive with decreasing development till the surface of astral objects. The graphical behavior of charge density can be seen in Fig. (\textbf{5}) against the different values of parameter $n$. Following Mafa et al. \cite{Takisa}, we initially tried to do the analysis with the rational form of electric field. However, we could not obtain physically realistic results in this case. Perhaps the complications arise due to the term involved $\frac{3Mq^2}{r^3}$ in Bardeen model corresponding to the the term $\frac{q^2}{r^2}$ in the usual Reissner Nordstrom model. Thus for the sake of simplicity, we have chosen linear form of electric field $E^2 =KQr$ which provides excellent results in our case.
It can be observed that the electric field is zero at $r=0$, and realize maximum near the boundary of stellar objects. The graphical analysis can be identified from the Fig. (\textbf{6}) for both cases.
\begin{figure}
\centering \epsfig{file=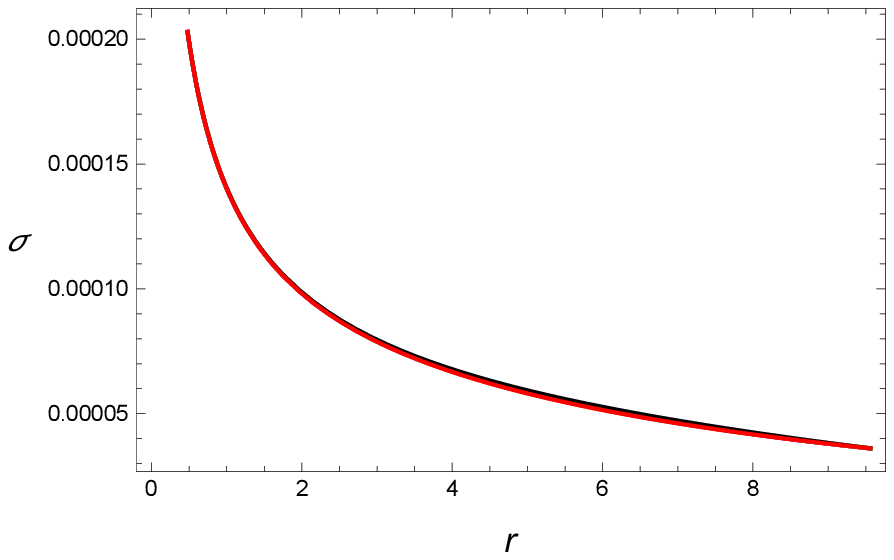, width=.48\linewidth,
height=2.1in}\epsfig{file=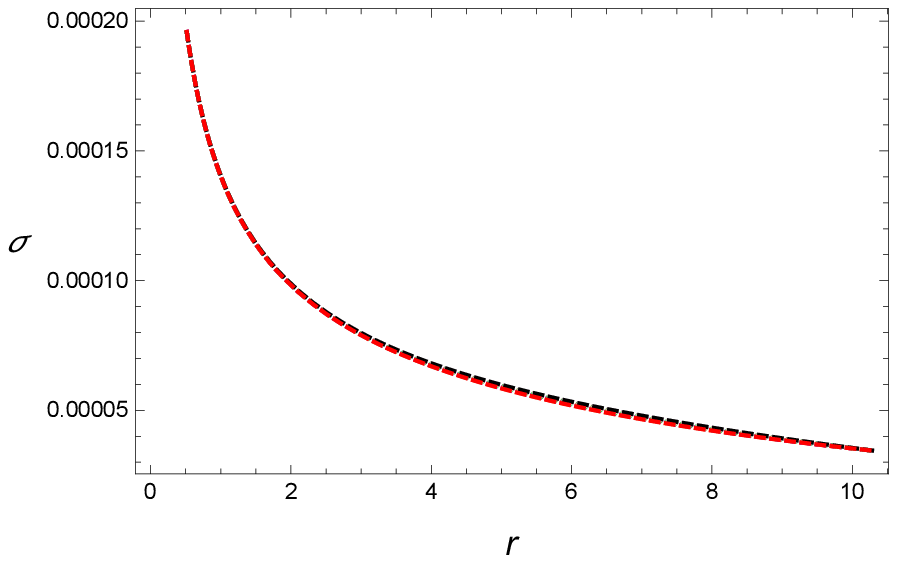, width=.48\linewidth,
height=2.1in}\caption{\label{Fig.5} describes the nature of charge density for two different models under $(M=1.77, \;R_{b}=9.56)$ and $(M=1.97, \;R_{b}=10.3)$ with $n=1.80(\textcolor{red}{\bigstar})$, $n=2.20(\textcolor{orange}{\bigstar})$, $n=2.60(\textcolor{blue}{\bigstar})$, $n=3.00(\textcolor{magenta}{\bigstar})$, $n=3.40(\textcolor{purple}{\bigstar})$, $n=3.80(\textcolor{pink}{\bigstar})$, $n=4.20(\textcolor{yellow}{\bigstar})$, $n=6.60(\textcolor{cyan}{\bigstar})$, $n=5.00(\textcolor{green}{\bigstar})$, $n=5.40(\textcolor{brown}{\bigstar})$, $n=5.80(\textcolor{gray}{\bigstar})$,
$n=6.20(\textcolor{orange!50}{\bigstar})$, $n=6.60(\textcolor{magenta!50}{\bigstar})$, and $n=7.00(\textcolor{black}{\bigstar})$.}
\end{figure}
\begin{figure}
\centering \epsfig{file=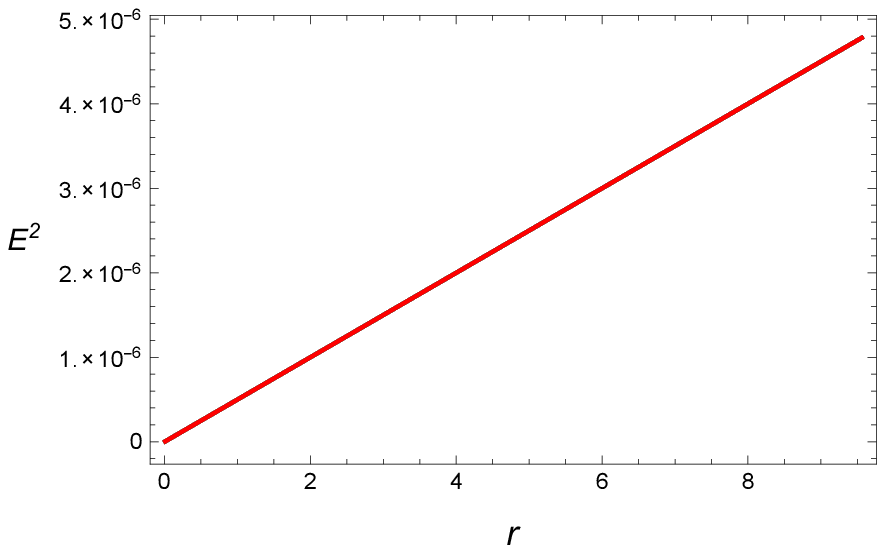, width=.48\linewidth,
height=2.1in}\epsfig{file=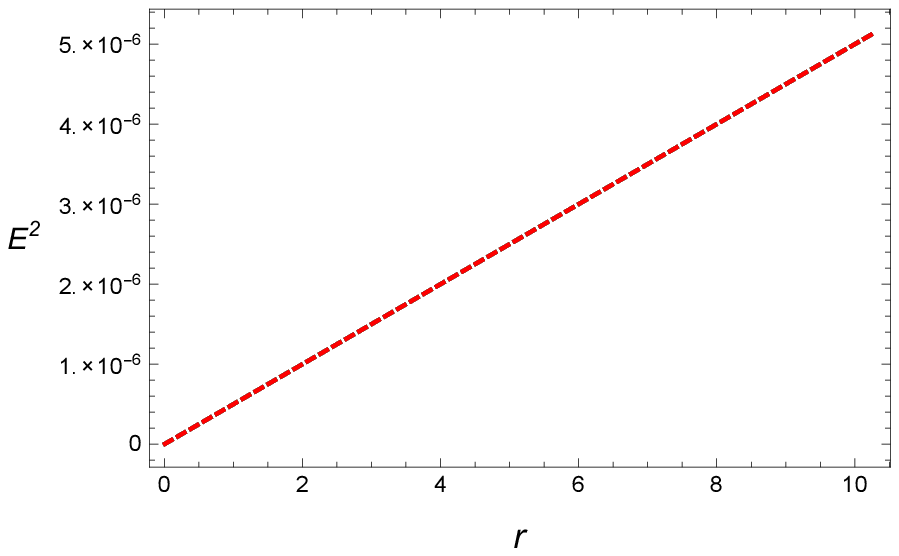, width=.48\linewidth,
height=2.1in} \caption{\label{Fig.6} describes the graphic nature of electric field for two different models under $(M=1.77, \;R_{b}=9.56)$ and $(M=1.97, \;R_{b}=10.3)$ with $n=1.80(\textcolor{red}{\bigstar})$, $n=2.20(\textcolor{orange}{\bigstar})$, $n=2.60(\textcolor{blue}{\bigstar})$, $n=3.00(\textcolor{magenta}{\bigstar})$, $n=3.40(\textcolor{purple}{\bigstar})$, $n=3.80(\textcolor{pink}{\bigstar})$, $n=4.20(\textcolor{yellow}{\bigstar})$, $n=6.60(\textcolor{cyan}{\bigstar})$, $n=5.00(\textcolor{green}{\bigstar})$, $n=5.40(\textcolor{brown}{\bigstar})$, $n=5.80(\textcolor{gray}{\bigstar})$,
$n=6.20(\textcolor{orange!50}{\bigstar})$, $n=6.60(\textcolor{magenta!50}{\bigstar})$, and $n=7.00(\textcolor{black}{\bigstar})$.}
\end{figure}
\subsubsection{Anisotropy}
The anisotropy profile using Eq. (\ref{new}) for the present study is calculated as:
\begin{eqnarray}
\triangle=p_{t}-p_{r}&=&\frac{r}{8 \pi  \left(Y r^2+1\right) \sqrt{F_1(r)} F_5(r)}\bigg(-\frac{B c r F(r) F_4(r) \left(X r^2+1\right)^n+Y n \left(Y r^2+1\right) F_3(r) F_2(r){}^{n/2}}{c r^2 \left(X r^2+1\right)^n+\left(Y r^2+1\right)^2}\nonumber\\&-&\frac{B c r F(r) F_7(r) \left(X r^2+1\right)^n+Y n \left(Y r^2+1\right) F_9(r) F_2(r){}^{n/2}}{\left(X r^2+1\right) F_6(r){}^2}\bigg).\label{34}
\end{eqnarray}
The Fig. (\textbf{7}) illustrates the behavior of anisotropy profile for two different models under the estimated values from Table. (\textbf{1}) and Table. (\textbf{2}), which is regarded as positive with increasing behavior near the boundary. The positive nature of $\triangle$ with electric charge shows the superiority of this study.
\begin{figure}
\centering \epsfig{file=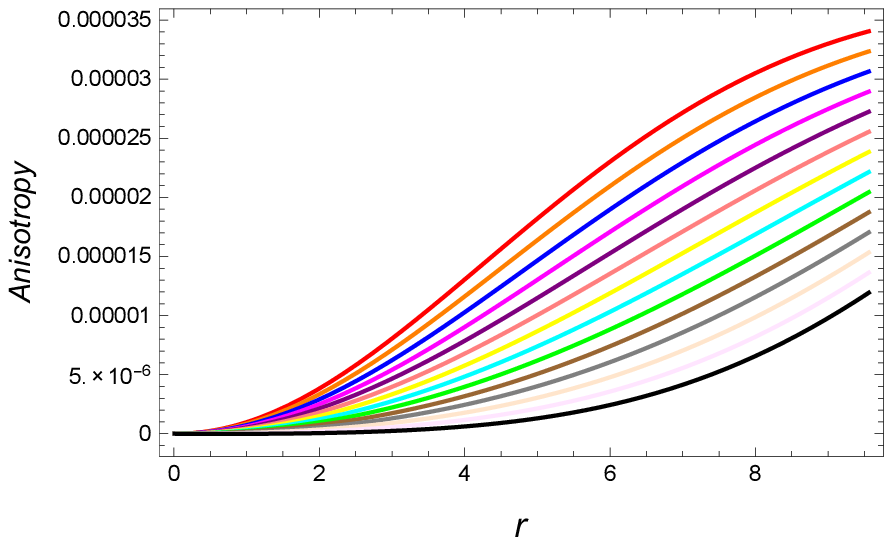, width=.48\linewidth,
height=2.1in}\epsfig{file=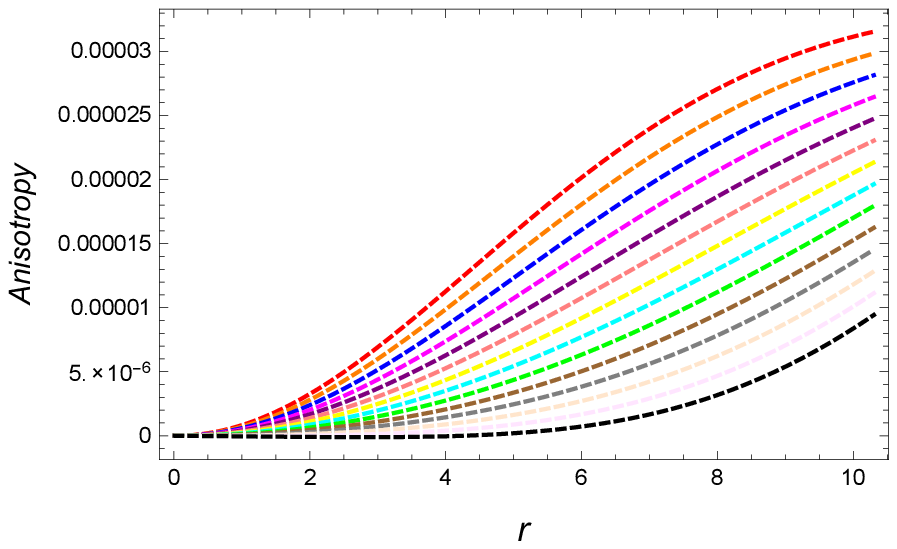, width=.48\linewidth,
height=2.1in} \caption{\label{Fig.7} describes the graphic nature of anisotropy for two different models under $(M=1.77, \;R_{b}=9.56)$ and $(M=1.97, \;R_{b}=10.3)$ with $n=1.80(\textcolor{red}{\bigstar})$, $n=2.20(\textcolor{orange}{\bigstar})$, $n=2.60(\textcolor{blue}{\bigstar})$, $n=3.00(\textcolor{magenta}{\bigstar})$, $n=3.40(\textcolor{purple}{\bigstar})$, $n=3.80(\textcolor{pink}{\bigstar})$, $n=4.20(\textcolor{yellow}{\bigstar})$, $n=6.60(\textcolor{cyan}{\bigstar})$, $n=5.00(\textcolor{green}{\bigstar})$, $n=5.40(\textcolor{brown}{\bigstar})$, $n=5.80(\textcolor{gray}{\bigstar})$,
$n=6.20(\textcolor{orange!50}{\bigstar})$, $n=6.60(\textcolor{magenta!50}{\bigstar})$, and $n=7.00(\textcolor{black}{\bigstar})$.}
\end{figure}
\subsubsection{Gradients}
This section is devoted to calculate the derivatives of the energy density function, and pressure components with respect to radial coordinate $r$. These derivatives are calculated as
\begin{eqnarray}
\frac{d\rho}{dr}&=&-\frac{\frac{2 X c (n-1) r F_0(r) \left(X r^2+1\right)^{n-2}}{D_5^2}-\frac{4 c D_4 r F_0(r) \left(X r^2+1\right)^{n-1}}{D_5^3}-\frac{2 c D_{13} r \left(X r^2+1\right)^{n-1}}{F_6(r)^2}+K Q}{8\pi},\label{35}\\
\frac{d p_{r}}{dr}&=&\frac{1}{8 \pi  D_5^2 \left(Y r^2+1\right)^2 \sqrt{F_1(r)} F_5(r)^2}\bigg(D_5 r \left(Y r^2+1\right) F_5(r) \left(B c D_2 r F(r) \left(X r^2+1\right)^n+2 X B c n r^2 F(r)\right.\nonumber\\ &\times&\left. F_4(r) \left(X r^2+1\right)^{n-1}+B c F(r) F_4(r) \left(X r^2+1\right)^n+2 Y^2 n r F_3(r) F_2(r){}^{n/2}+Y \left(D_8+D_9\right) n \left(Y r^2+1\right)\right.\nonumber\\ &\times&\left. F_2(r)^{n/2}-D_6+D_{10}\right)-\frac{D_5 D_1 D_3 F_5(r)}{X r^2+1}-2 D_4 D_3 r^2 \left(Y r^2+1\right) F_5(r)-2 Y D_5 D_3 r^2 F_5(r)+D_3 D_5\nonumber\\ &\times& \left(Y r^2+1\right) F_5(r)-D_5 \left(D_{11}+D_{12}\right) D_3 r \left(Y r^2+1\right)\bigg),\label{36}\\
\frac{d p_{t}}{dr}&=&\frac{1}{8 \pi  \left(X r^2+1\right)^2 \left(Y r^2+1\right)^2 \sqrt{F_1(r)} F_6(r){}^3 F_5(r){}^2}\bigg(-r \left(X r^2+1\right) \left(Y r^2+1\right) F_6(r) F_5(r) \left(D_{16} F_2(r){}^{n/2} \right.\nonumber\\ &\times&\left. \left(-2 B c D_{21} r^2 \left(X r^2+1\right)^n+A \left(D_{18}+D_{19}+D_{20}\right) \sqrt{F_1(r)}+D_{17}\right)+B c D_{23} r F(r) \left(X r^2+1\right)^n\right.\nonumber\\ &+&\left. 2 X B c n r^2 F(r) F_7(r) \left(X r^2+1\right)^{n-1}+B c F(r) F_7(r) \left(X r^2+1\right)^n+D_{22}\right)+4 D_4 D_7 r^2 \left(X r^2+1\right)\nonumber\\ &\times& \left(Y r^2+1\right) F_5(r)+2 Y D_7 r^2 \left(X r^2+1\right)F_5(r) F_6(r)+2 X D_7 r^2 \left(Y r^2+1\right) F_5(r) F_6(r)+D_7 \left(D_{14}+D_{15}\right)\nonumber\\ &\times& r \left(X r^2+1\right) \left(Y r^2+1\right) F_6(r)-D_7 \left(X r^2+1\right) \left(Y r^2+1\right) F_6(r) F_5(r)+D_1 D_7 F_5(r) F_6(r)\bigg),\label{37}
\end{eqnarray}
where $D_i$, $\{i=1,...,23\}$ are given in the Appendix (\textbf{II}).
From the Fig. (\textbf{8}), it is reported that the gradients, i.e., $\frac{d\rho}{dr},\;\frac{dp_{r}}{dr},\;\& \frac{dp_{t}}{dr}$ are examined negative for all the different values of parameter $n$. The negative behavior of gradients shows the supremacy and perfection of this model with electric charge.
\begin{figure}
\centering \epsfig{file=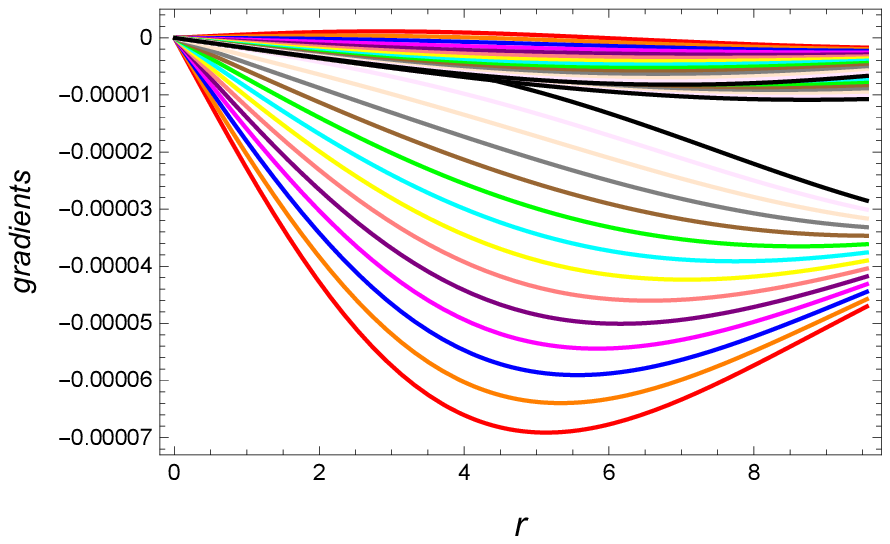, width=.48\linewidth,
height=2.1in}\epsfig{file=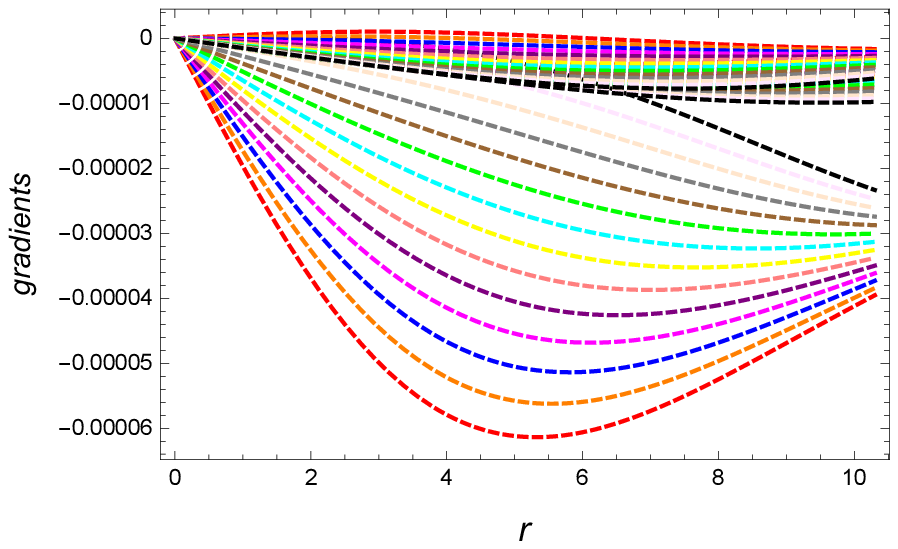, width=.48\linewidth,
height=2.1in} \caption{\label{Fig.8} describes the graphic nature of gardients for two different models under $(M=1.77, \;R_{b}=9.56)$ and $(M=1.97, \;R_{b}=10.3)$ with $n=1.80(\textcolor{red}{\bigstar})$, $n=2.20(\textcolor{orange}{\bigstar})$, $n=2.60(\textcolor{blue}{\bigstar})$, $n=3.00(\textcolor{magenta}{\bigstar})$, $n=3.40(\textcolor{purple}{\bigstar})$, $n=3.80(\textcolor{pink}{\bigstar})$, $n=4.20(\textcolor{yellow}{\bigstar})$, $n=6.60(\textcolor{cyan}{\bigstar})$, $n=5.00(\textcolor{green}{\bigstar})$, $n=5.40(\textcolor{brown}{\bigstar})$, $n=5.80(\textcolor{gray}{\bigstar})$,
$n=6.20(\textcolor{orange!50}{\bigstar})$, $n=6.60(\textcolor{magenta!50}{\bigstar})$, and $n=7.00(\textcolor{black}{\bigstar})$.}
\end{figure}
\begin{center}
\begin{table}
\caption{\label{tab1}{Summary of EoS parameters, equilibrium forces, energy conditions, and mass, compactness and red-shift functions.}}
\begin{tabular}{|c|c|c|c|c|c|c|c|c|}
    \hline
\multicolumn{4}{|c|}{$(M=1.77, \;R_{b}=9.56)\;\&\;(M=1.97, \;R_{b}=10.3)$}\\
    \hline
$n$        & $\triangle$                           &$\frac{d\rho}{dr}\&\frac{dp_{r}}{dr}\& \frac{dp_{t}}{dr}$                         & $\frac{p_{rc}}{\rho_{c}}=\frac{p_{tc}}{\rho_{c}}$(Zeldovich's condition)\\
\hline
1.80     & $\triangle>0$                         &$\frac{d\rho}{dr}<0\&\frac{dp_{r}}{dr}<0\& \frac{dp_{t}}{dr}>0$           &$\frac{p_{rc}}{\rho_{c}}=\frac{p_{tc}}{\rho_{c}}<1$\\

2.20    & $\triangle>0$                         &$\frac{d\rho}{dr}<0\&\frac{dp_{r}}{dr}<0\& \frac{dp_{t}}{dr}>0$           &$\frac{p_{rc}}{\rho_{c}}=\frac{p_{tc}}{\rho_{c}}<1$\\

2.60   & $\triangle>0$                         &$\frac{d\rho}{dr}<0\&\frac{dp_{r}}{dr}<0\& \frac{dp_{t}}{dr}<0$           &$\frac{p_{rc}}{\rho_{c}}=\frac{p_{tc}}{\rho_{c}}<1$\\

3.00    & $\triangle>0$                         &$\frac{d\rho}{dr}<0\&\frac{dp_{r}}{dr}<0\& \frac{dp_{t}}{dr}<0$           &$\frac{p_{rc}}{\rho_{c}}=\frac{p_{tc}}{\rho_{c}}<1$\\

3.40   & $\triangle>0$                         &$\frac{d\rho}{dr}<0\&\frac{dp_{r}}{dr}<0\& \frac{dp_{t}}{dr}<0$           &$\frac{p_{rc}}{\rho_{c}}=\frac{p_{tc}}{\rho_{c}}<1$\\

3.80   & $\triangle>0$                         &$\frac{d\rho}{dr}<0\&\frac{dp_{r}}{dr}<0\& \frac{dp_{t}}{dr}<0$           &$\frac{p_{rc}}{\rho_{c}}=\frac{p_{tc}}{\rho_{c}}<1$\\

4.20   & $\triangle>0$                         &$\frac{d\rho}{dr}<0\&\frac{dp_{r}}{dr}<0\& \frac{dp_{t}}{dr}<0$           &$\frac{p_{rc}}{\rho_{c}}=\frac{p_{tc}}{\rho_{c}}<1$\\

4.60   & $\triangle>0$                         &$\frac{d\rho}{dr}<0\&\frac{dp_{r}}{dr}<0\& \frac{dp_{t}}{dr}<0$           &$\frac{p_{rc}}{\rho_{c}}=\frac{p_{tc}}{\rho_{c}}<1$\\

5.00  & $\triangle>0$                         &$\frac{d\rho}{dr}<0\&\frac{dp_{r}}{dr}<0\& \frac{dp_{t}}{dr}<0$           &$\frac{p_{rc}}{\rho_{c}}=\frac{p_{tc}}{\rho_{c}}<1$\\

5.40    & $\triangle>0$                         &$\frac{d\rho}{dr}<0\&\frac{dp_{r}}{dr}<0\& \frac{dp_{t}}{dr}<0$           &$\frac{p_{rc}}{\rho_{c}}=\frac{p_{tc}}{\rho_{c}}<1$\\

5.80    & $\triangle>0$                         &$\frac{d\rho}{dr}<0\&\frac{dp_{r}}{dr}<0\& \frac{dp_{t}}{dr}<0$           &$\frac{p_{rc}}{\rho_{c}}=\frac{p_{tc}}{\rho_{c}}<1$\\

6.20    & $\triangle>0$                         &$\frac{d\rho}{dr}<0\&\frac{dp_{r}}{dr}<0\& \frac{dp_{t}}{dr}<0$           &$\frac{p_{rc}}{\rho_{c}}=\frac{p_{tc}}{\rho_{c}}<1$\\

6.60   & $\triangle>0$                         &$\frac{d\rho}{dr}<0\&\frac{dp_{r}}{dr}<0\& \frac{dp_{t}}{dr}<0$           &$\frac{p_{rc}}{\rho_{c}}=\frac{p_{tc}}{\rho_{c}}<1$\\

7.00   & $\triangle>0$                         &$\frac{d\rho}{dr}<0\&\frac{dp_{r}}{dr}<0\& \frac{dp_{t}}{dr}<0$           &$\frac{p_{rc}}{\rho_{c}}=\frac{p_{tc}}{\rho_{c}}<1$\\
\hline
\end{tabular}
\end{table}
\end{center}
\subsubsection{Energy Conditions}
In this section, we shall explore the basic aspiration of energy bounds in the scope of GR under electric filed for this current study. The energy bounds are considered as the main part of compact stars study. There are four kinds of energy bounds, which are described as
\begin{eqnarray}
\rho \geq 0&&\;\;\;\;\Rightarrow \;\;\;\; \;\;\;\; NEC\label{38}\\
\rho-p_{t}\geq0,\;\;\;\;\rho-p_{r}\geq0&&\;\;\;\;\Rightarrow \;\;\;\;\;\;\;\;  WEC\label{39}\\
\rho-p_{r}-2p_{t}\geq 0&&\;\;\;\;\Rightarrow \;\;\;\;\;\;\;\;  SEC\label{40}\\
\rho >|p_r|,\;\;\;\;\rho >|p_t|&&\;\;\;\;\Rightarrow \;\;\;\;\;\;\;\;  DEC\label{41}
\end{eqnarray}
\begin{itemize}
  \item $(NEC)$ denotes the null energy condition,
  \item $(NEC)$ defines the weak energy condition,
  \item $(SEC)$ represents the strong energy condition,
  \item $(DEC)$ illustrates the dominant energy condition,
\end{itemize}

Here, in this study all the energy bounds are seen satisfied. As evident from Fig. (\textbf{1}), the $NEC$ is satisfied. The Fig. (\textbf{9}) shows the presence of remaining energy bounds, which are noticed satisfied, i.e., all the energy conditions remain positive thorough out this study. The validity of energy conditions shows the preeminence and dominance of this model.
\begin{figure}
\centering \epsfig{file=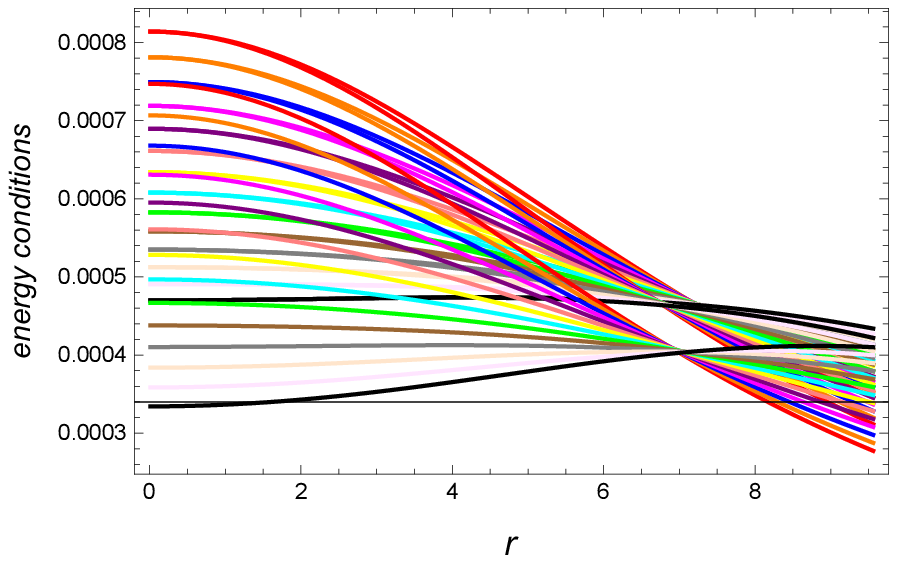, width=.48\linewidth,
height=2.1in}\epsfig{file=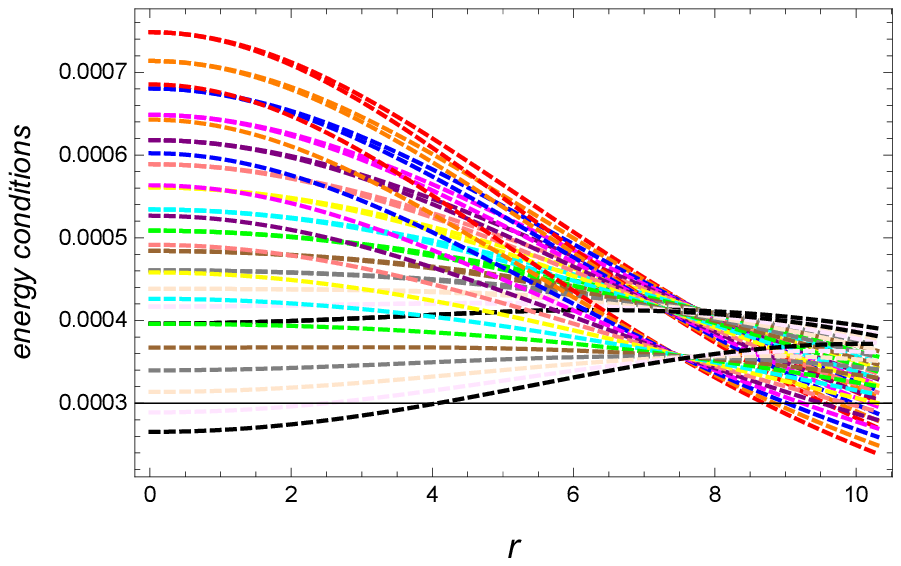, width=.48\linewidth,
height=2.1in} \caption{\label{Fig.9} describes the positive behavior of energy conditions for two different models under $(M=1.77, \;R_{b}=9.56)$ and $(M=1.97, \;R_{b}=10.3)$ with $n=1.80(\textcolor{red}{\bigstar})$, $n=2.20(\textcolor{orange}{\bigstar})$, $n=2.60(\textcolor{blue}{\bigstar})$, $n=3.00(\textcolor{magenta}{\bigstar})$, $n=3.40(\textcolor{purple}{\bigstar})$, $n=3.80(\textcolor{pink}{\bigstar})$, $n=4.20(\textcolor{yellow}{\bigstar})$, $n=6.60(\textcolor{cyan}{\bigstar})$, $n=5.00(\textcolor{green}{\bigstar})$, $n=5.40(\textcolor{brown}{\bigstar})$, $n=5.80(\textcolor{gray}{\bigstar})$,
$n=6.20(\textcolor{orange!50}{\bigstar})$, $n=6.60(\textcolor{magenta!50}{\bigstar})$, and $n=7.00(\textcolor{black}{\bigstar})$.}
\end{figure}
\subsubsection{Equation of state parameters}
Two famous ratios between radial and tangential pressure sources and energy density, i.e., $\frac{p_r}{\rho}$ and $\frac{p_t}{\rho}$ explain the concept of EoS parameters, i.e., $w_r$ and $w_t$, which are calculated as
\begin{eqnarray}
w_r =\frac{p_r}{\rho}&&=-\frac{r \left(B c r F(r) F_4(r) \left(X r^2+1\right)^n+Y n \left(Y r^2+1\right) F_3(r) F_2(r){}^{n/2}\right)}{w_1 \left(Y r^2+1\right) \sqrt{F_1(r)} F_5(r) \left(c r^2 \left(X r^2+1\right)^n+\left(Y r^2+1\right)^2\right)},\label{42}\\
w_t =\frac{p_t}{\rho}&&=\frac{r \left(B c r w_3 F(r) \left(X r^2+1\right)^n+Y n w_2 \left(Y r^2+1\right) F_2(r){}^{n/2}\right)}{w_1 \left(X r^2+1\right) \left(Y r^2+1\right) \sqrt{F_1(r)} F_5(r) \left(c r^2 \left(X r^2+1\right)^n+\left(Y r^2+1\right)^2\right)^2},\label{43}
\end{eqnarray}
where $w_i$, $\{i=1,2,3\}$ are given in the Appendix (\textbf{III}). It is worthwhile to mention here that the constraints $w_i$ play an important role in exploring the nature of stellar structures. In our case, we may obtain $-1<\omega_r<-1/3$ for fixing the suitable values of $w_1$. Thus Bardeen stellar structures with Karmarkar Condition may support so-called dark models of stellar structures \cite{20Bhar5}-\cite{22Rahaman4}. However, a detailed analysis is left for future work.
\begin{figure}
\centering \epsfig{file=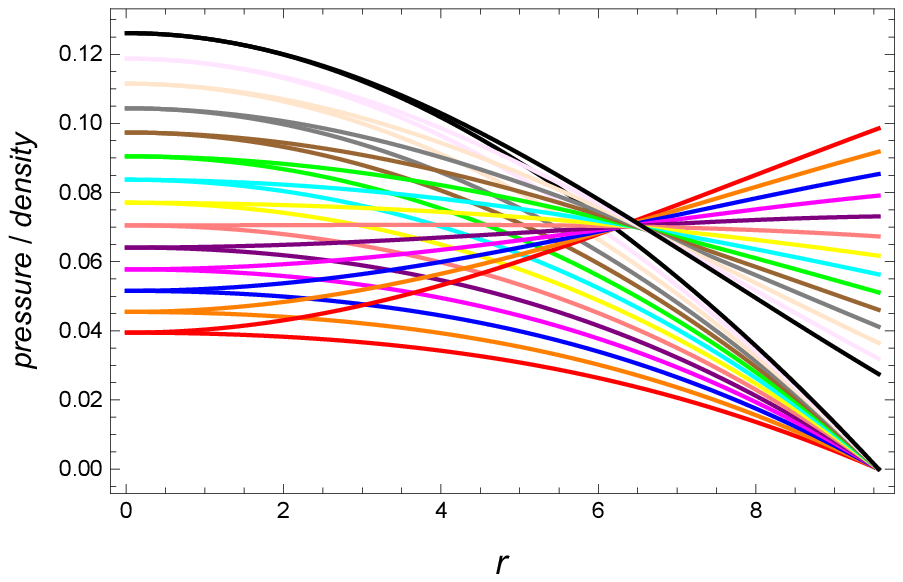, width=.48\linewidth,
height=2.1in}\epsfig{file=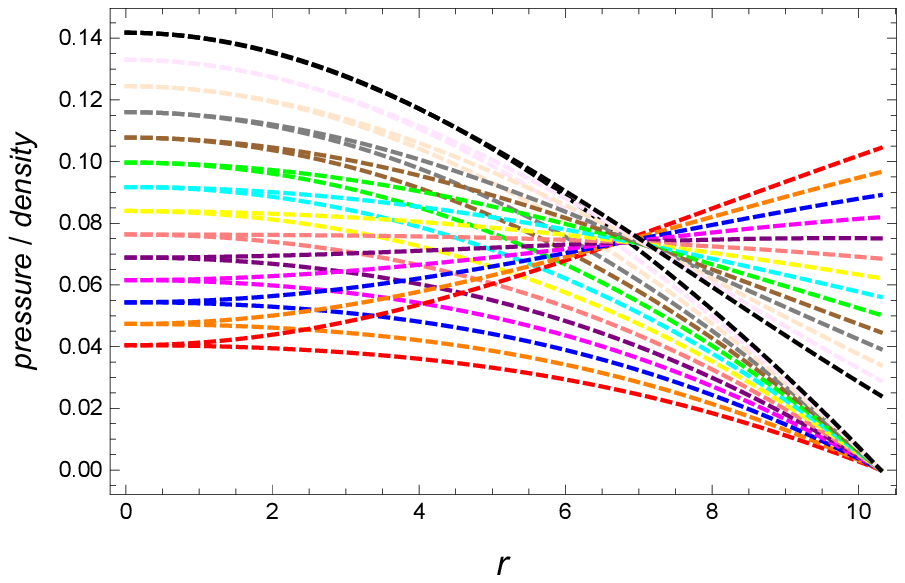, width=.48\linewidth,
height=2.1in} \caption{\label{Fig.10} shows the equation od state parameters behavior of two different models under $(M=1.77, \;R_{b}=9.56)$ and $(M=1.97, \;R_{b}=10.3)$ with $n=1.80(\textcolor{red}{\bigstar})$, $n=2.20(\textcolor{orange}{\bigstar})$, $n=2.60(\textcolor{blue}{\bigstar})$, $n=3.00(\textcolor{magenta}{\bigstar})$, $n=3.40(\textcolor{purple}{\bigstar})$, $n=3.80(\textcolor{pink}{\bigstar})$, $n=4.20(\textcolor{yellow}{\bigstar})$, $n=6.60(\textcolor{cyan}{\bigstar})$, $n=5.00(\textcolor{green}{\bigstar})$, $n=5.40(\textcolor{brown}{\bigstar})$, $n=5.80(\textcolor{gray}{\bigstar})$,
$n=6.20(\textcolor{orange!50}{\bigstar})$, $n=6.60(\textcolor{magenta!50}{\bigstar})$, and $n=7.00(\textcolor{black}{\bigstar})$.}
\end{figure}

Now, we check the graphical representation of equations of state parameters, i.e., $w_r\;\&\;w_t$. It is cleared from the Fig. (\textbf{10}), that both the radial and tangential parameters are remained positive and also reported less than 1, i.e., $0\leq w_r\;\&\;w_t<1$ for both models. The radial EoS parameter is observed zero at boundary. The tangential parameter remains greater than zero throughout the configurations. The condition $0\leq w_r\;\&\;w_t<1$ shows that our results are correct with excellency in accuracy.

\subsubsection{Mass-function, compactness parameter, and red-shift function}
The mass-function, compactness parameter, and red-shift functions are considered the necessary components for the stellar study. All these physical parameters are linked to each others. The mass-function is calculated as
\begin{equation}\label{44}
m(r)=4\pi \int^{r}_{0}(\rho \times r^{2})dr=\frac{1}{2} \left(-\frac{1}{\frac{c r \left(X r^2+1\right)^n}{\left(Y r^2+1\right)^2}+\frac{1}{r}}-\frac{1}{4} K Q r^4+r\right),
\end{equation}
where $m(r)$ represents the mass-function. The compactness parameter is denoted by $u$, and it is calculated as
\begin{equation}\label{45}
u=\frac{2}{r}m(r)=\frac{1}{r}  \left(-\frac{1}{\frac{c r \left(X r^2+1\right)^n}{\left(Y r^2+1\right)^2}+\frac{1}{r}}-\frac{1}{4} K Q r^4+r\right),\\
\end{equation}
The red-shift function is discussed for this current study as
\begin{equation}\label{46}
Z_{s}=(1-2 u)^{-\frac{1}{2}}-1=\left(1- \frac{2}{r}  \left(-\frac{1}{\frac{c r \left(X r^2+1\right)^n}{\left(Y r^2+1\right)^2}+\frac{1}{r}}-\frac{1}{4} K Q r^4+r\right)\right)^{-\frac{1}{2}}-1.
\end{equation}
where $u$ denotes the compactness parameter.
\begin{figure}
\centering \epsfig{file=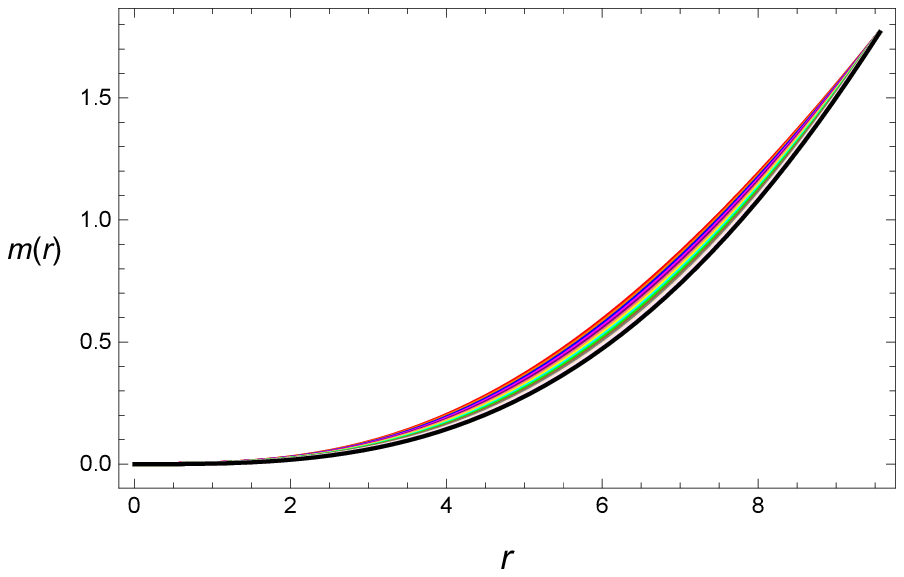, width=.48\linewidth,
height=2.1in}\epsfig{file=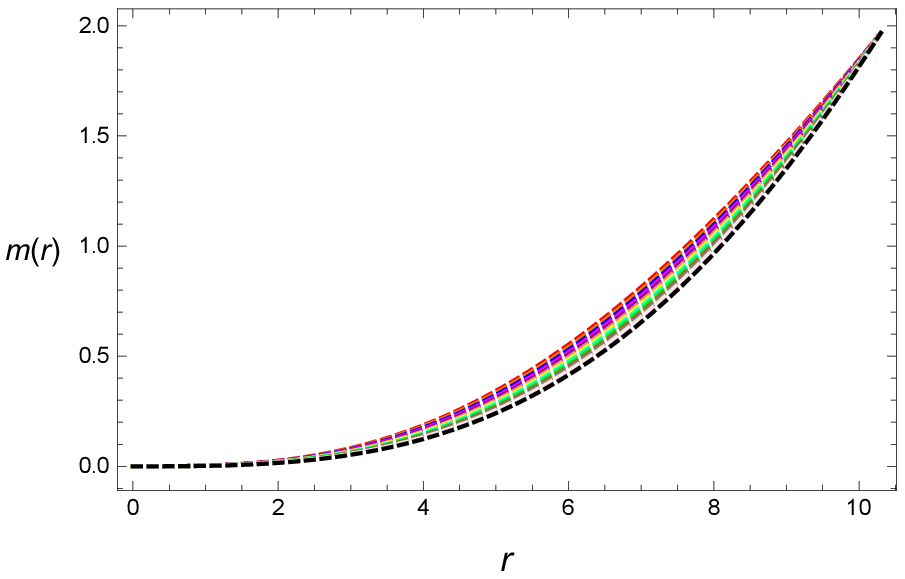, width=.48\linewidth,
height=2.1in} \caption{\label{Fig.11} describes the graphic nature of mass-function for two different models under $(M=1.77, \;R_{b}=9.56)$ and $(M=1.97, \;R_{b}=10.3)$ with $n=1.80(\textcolor{red}{\bigstar})$, $n=2.20(\textcolor{orange}{\bigstar})$, $n=2.60(\textcolor{blue}{\bigstar})$, $n=3.00(\textcolor{magenta}{\bigstar})$, $n=3.40(\textcolor{purple}{\bigstar})$, $n=3.80(\textcolor{pink}{\bigstar})$, $n=4.20(\textcolor{yellow}{\bigstar})$, $n=6.60(\textcolor{cyan}{\bigstar})$, $n=5.00(\textcolor{green}{\bigstar})$, $n=5.40(\textcolor{brown}{\bigstar})$, $n=5.80(\textcolor{gray}{\bigstar})$,
$n=6.20(\textcolor{orange!50}{\bigstar})$, $n=6.60(\textcolor{magenta!50}{\bigstar})$, and $n=7.00(\textcolor{black}{\bigstar})$.}
\end{figure}
\begin{figure}
\centering \epsfig{file=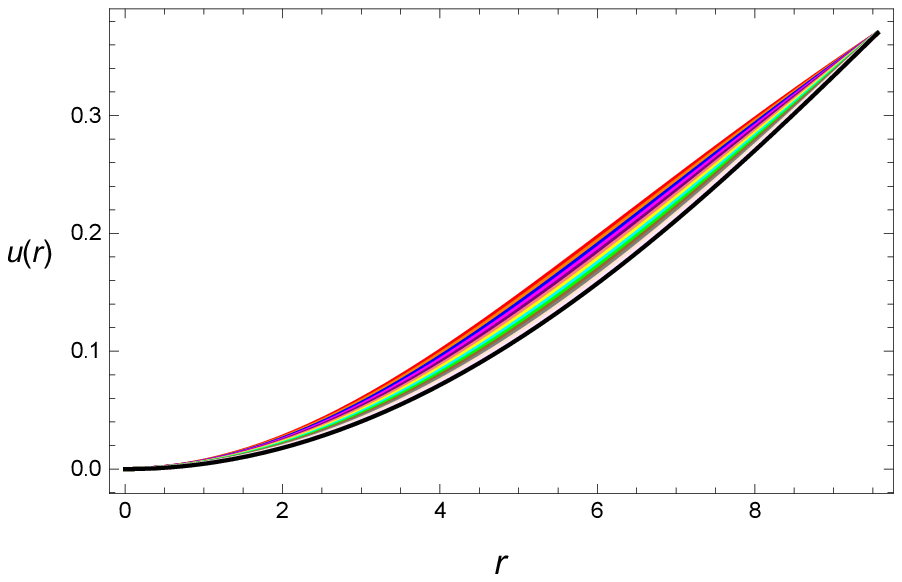, width=.48\linewidth,
height=2.1in}\epsfig{file=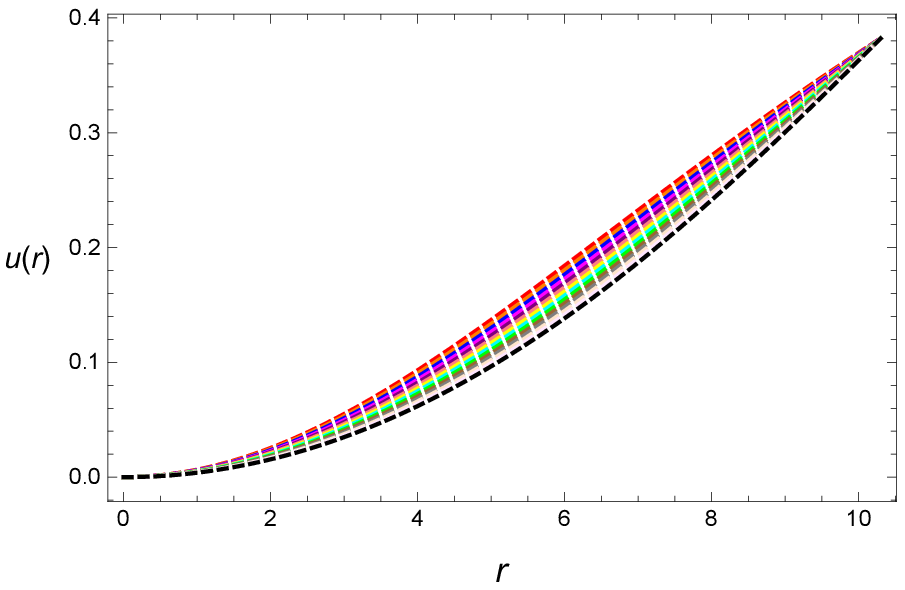, width=.48\linewidth,
height=2.1in} \caption{\label{Fig.12} discusses the development of compactness parameter for two different models under $(M=1.77, \;R_{b}=9.56)$ and $(M=1.97, \;R_{b}=10.3)$ with $n=1.80(\textcolor{red}{\bigstar})$, $n=2.20(\textcolor{orange}{\bigstar})$, $n=2.60(\textcolor{blue}{\bigstar})$, $n=3.00(\textcolor{magenta}{\bigstar})$, $n=3.40(\textcolor{purple}{\bigstar})$, $n=3.80(\textcolor{pink}{\bigstar})$, $n=4.20(\textcolor{yellow}{\bigstar})$, $n=6.60(\textcolor{cyan}{\bigstar})$, $n=5.00(\textcolor{green}{\bigstar})$, $n=5.40(\textcolor{brown}{\bigstar})$, $n=5.80(\textcolor{gray}{\bigstar})$,
$n=6.20(\textcolor{orange!50}{\bigstar})$, $n=6.60(\textcolor{magenta!50}{\bigstar})$, and $n=7.00(\textcolor{black}{\bigstar})$. }
\end{figure}
\begin{figure}
\centering \epsfig{file=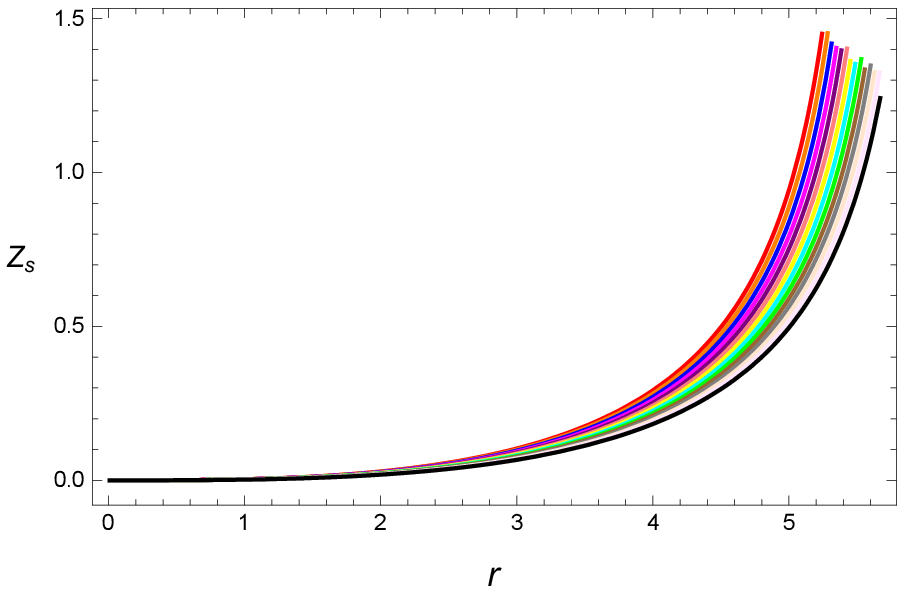, width=.48\linewidth,
height=2.1in}\epsfig{file=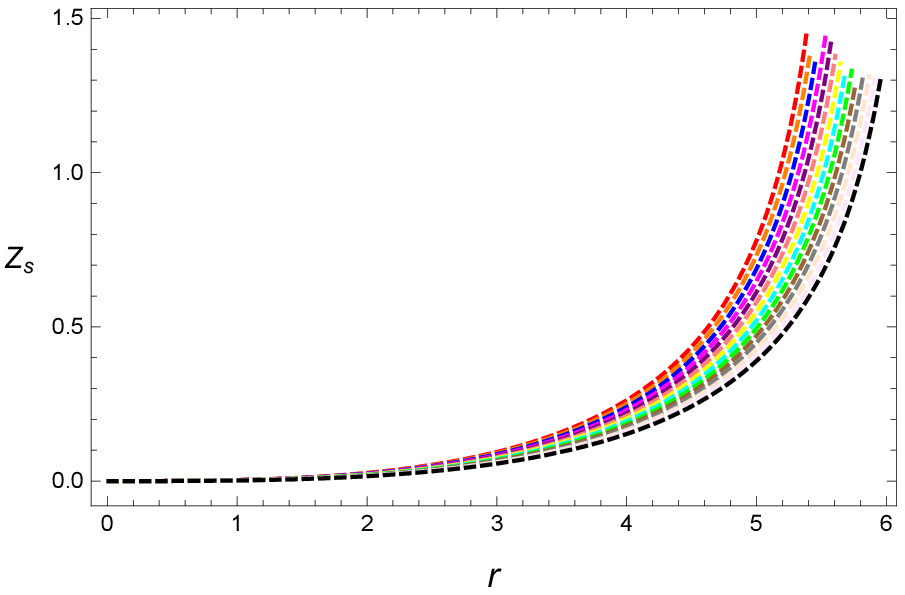, width=.48\linewidth,
height=2.1in} \caption{\label{Fig.13} describe the graphic nature of red-shift function for two different models under $(M=1.77, \;R_{b}=9.56)$ and $(M=1.97, \;R_{b}=10.3)$ with $n=1.80(\textcolor{red}{\bigstar})$, $n=2.20(\textcolor{orange}{\bigstar})$, $n=2.60(\textcolor{blue}{\bigstar})$, $n=3.00(\textcolor{magenta}{\bigstar})$, $n=3.40(\textcolor{purple}{\bigstar})$, $n=3.80(\textcolor{pink}{\bigstar})$, $n=4.20(\textcolor{yellow}{\bigstar})$, $n=6.60(\textcolor{cyan}{\bigstar})$, $n=5.00(\textcolor{green}{\bigstar})$, $n=5.40(\textcolor{brown}{\bigstar})$, $n=5.80(\textcolor{gray}{\bigstar})$,
$n=6.20(\textcolor{orange!50}{\bigstar})$, $n=6.60(\textcolor{magenta!50}{\bigstar})$, and $n=7.00(\textcolor{black}{\bigstar})$.}
\end{figure}
With the help of above three equations, we discuss the graphically representations of these three different parameters like mass-function, compactness, and red-shift function, i.e., $m(r),\;u(r),\;\&~Z_{s}$. The mass-function is seen very closer to observed mass for both the models, i.e., $(M=1.77, \;R_{b}=9.56)$ and $(M=1.97, \;R_{b}=10.3)$, it can verified from the Fig. (\textbf{11}). The mass-function is realized from left penal of Fig. (\textbf{11}) that $m(r)\equiv 1.77\equiv M$ at boundary, and it is also observed from right penal that $m(r)\equiv 1.97\equiv M$ at boundary. The compactness parameter meets the Bhuchdahl \cite{Buchdahl} limit, i.e., $u<\frac{8}{9}$ in the current study, it can be recognized from the Fig. (\textbf{12}). The red shift function also satisfies the Mak and Harko \cite{MaK} condition. Further, the red-shift function also fulfills the Bohmer and Harko \cite{40} condition, i.e., $Z_{s}\leq 5$ and also it meets the Ivanov \cite{41} condition, i.e., $Z_{s}\leq 5.211$. Furthermore, the graphical representation of red-shift function can be revealed from the Fig. (\textbf{13}).
\begin{center}
\begin{table}
\caption{\label{tab1}{Summary of EoS parameters, equilibrium forces, energy conditions, and mass, compactness and red-shift functions.}}
\begin{tabular}{|c|c|c|c|c|c|c|c|c|}
    \hline
\multicolumn{5}{|c|}{$(M=1.77, \;R_{b}=9.56)\;\&\;(M=1.97, \;R_{b}=10.3)$}\\
    \hline
$n$    & $w_r\&w_t$           &$NEC\&WEC\&SEC\&DEC$  &$m(r)\&u(r)\&Z_s$        &$\mathscr{F}_{\mathrm{a}}\&\mathscr{F}_{\mathrm{h}}\&\mathscr{F}_{g}\& \mathscr{F}_{\mathrm{e}}$\\
\hline
1.80     & $0\leq w_r\&w_t<1$   &All are satisfied    &$m(r)>0\&u(r)< 8/9\&Z_s< 1.5$    &Balanced under TOV equation \\

2.20    & $0\leq w_r\&w_t<1$   &All are satisfied    &$m(r)>0\&u(r)< 8/9\&Z_s< 1.5$    &Balanced under TOV equation \\

2.60    & $0\leq w_r\&w_t<1$   &All are satisfied    &$m(r)>0\&u(r)< 8/9\&Z_s< 1.5$    &Balanced under TOV equation \\

3.00    & $0\leq w_r\&w_t<1$   &All are satisfied    &$m(r)>0\&u(r)< 8/9\&Z_s< 1.5$    &Balanced under TOV equation \\

3.40    & $0\leq w_r\&w_t<1$   &All are satisfied    &$m(r)>0\&u(r)< 8/9\&Z_s< 1.5$    &Balanced under TOV equation \\

3.80    & $0\leq w_r\&w_t<1$   &All are satisfied    &$m(r)>0\&u(r)< 8/9\&Z_s< 1.5$    &Balanced under TOV equation \\

4.20     & $0\leq w_r\&w_t<1$   &All are satisfied    &$m(r)>0\&u(r)< 8/9\&Z_s< 1.5$    &Balanced under TOV equation \\

4.60    & $0\leq w_r\&w_t<1$   &All are satisfied    &$m(r)>0\&u(r)< 8/9\&Z_s< 1.5$    &Balanced under TOV equation \\

5.00    & $0\leq w_r\&w_t<1$   &All are satisfied    &$m(r)>0\&u(r)< 8/9\&Z_s< 1.5$    &Balanced under TOV equation \\

5.40    & $0\leq w_r\&w_t<1$   &All are satisfied    &$m(r)>0\&u(r)< 8/9\&Z_s< 1.5$    &Balanced under TOV equation \\

5.80     & $0\leq w_r\&w_t<1$   &All are satisfied    &$m(r)>0\&u(r)< 8/9\&Z_s< 1.5$    &Balanced under TOV equation \\

6.20    & $0\leq w_r\&w_t<1$   &All are satisfied    &$m(r)>0\&u(r)< 8/9\&Z_s< 1.5$    &Balanced under TOV equation \\

6.60   & $0\leq w_r\&w_t<1$   &All are satisfied    &$m(r)>0\&u(r)< 8/9\&Z_s< 1.5$    &Balanced under TOV equation \\

7.00    & $0\leq w_r\&w_t<1$   &All are satisfied    &$m(r)>0\&u(r)< 8/9\&Z_s< 1.5$    &Balanced under TOV equation \\
\hline
\end{tabular}
\end{table}
\end{center}
\subsubsection{Equilibrium analysis}
The equilibrium analysis is a compulsory part of stellar study. Here, we are using the Bardeen geometry with electric charge to calculate the matching conditions. In this regard, we shall discuss the TOV equation with electric charge, which is defined as:
\begin{equation}\label{47}
-\sigma(r)E(r)e^{\frac{\lambda (r)}{2}}-\frac{2}{r}(p_{r}-p_{t})-\frac{d p _{r}}{dr}-\frac{\nu(r)'}{2}(\rho+p_{r})=0,
\end{equation}
The above equation can be summarized as:
\begin{equation}\label{48}
\mathscr{F}_{\mathrm{h}}+\mathscr{F}_{g}+\mathscr{F}_{\mathrm{a}}+\mathscr{F}_{\mathrm{e}}=0,
\end{equation}
where
\begin{eqnarray*}
\mathscr{F}_{\mathrm{a}}&&=\frac{2}{r}(p_{t}-p_{r})\;\;\;\;\;\;\;\;\;\;\;\;\Rightarrow \;\;\;\; \;\;\;\; AF,\\
\mathscr{F}_{\mathrm{h}}&&=-\frac{d p _{r}}{dr}\;\;\;\;\;\;\;\;\;\;\;\;\;\;\;\;\;\;\;\;\Rightarrow \;\;\;\; \;\;\;\;HF,\\
\mathscr{F}_{g}&&=-\frac{\nu(r)'}{2}(\rho+p_{r})\;\;\;\;\;\Rightarrow \;\;\;\; \;\;\;\;GF,\\
\mathscr{F}_{\mathrm{e}}&&=-\sigma(r)E(r)e^{\frac{\lambda (r)}{2}} \;\;\;\;\;\Rightarrow \;\;\;\; \;\;\;\;EF,
\end{eqnarray*}
where $AF$ represents the anisotropic force, $HF$ denotes the hydrostatic force, $GF$ mentions the gravitational force, and $EF$ presents the electric force.
\begin{figure}
\centering \epsfig{file=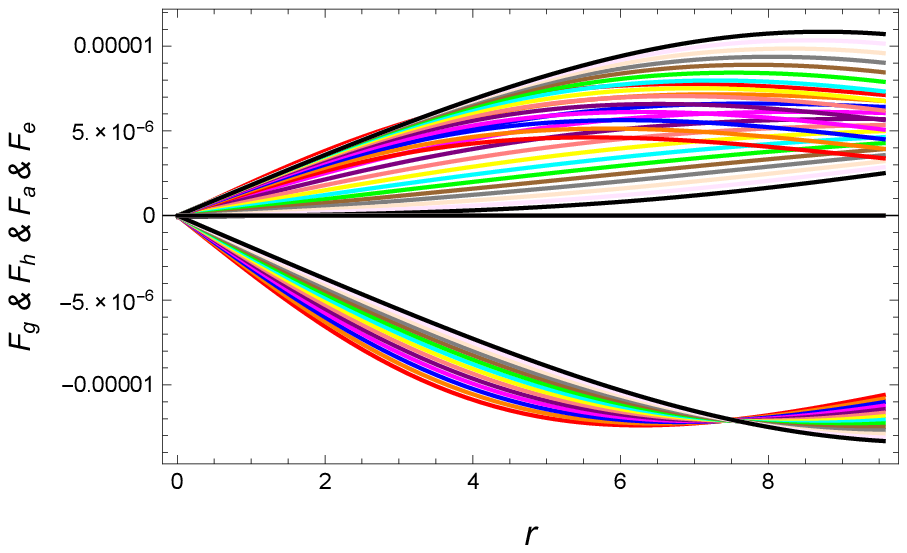, width=.48\linewidth,
height=2.1in}\epsfig{file=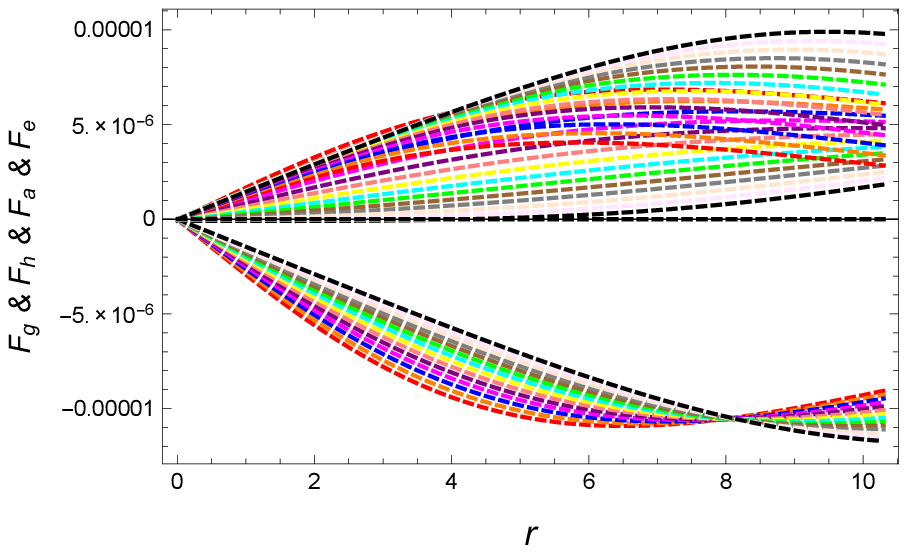, width=.48\linewidth,
height=2.1in} \caption{\label{Fig.14} shows the balancing nature of TOV equation for two different models under $(M=1.77, \;R_{b}=9.56)$ and $(M=1.97, \;R_{b}=10.3)$ with $n=1.80(\textcolor{red}{\bigstar})$, $n=2.20(\textcolor{orange}{\bigstar})$, $n=2.60(\textcolor{blue}{\bigstar})$, $n=3.00(\textcolor{magenta}{\bigstar})$, $n=3.40(\textcolor{purple}{\bigstar})$, $n=3.80(\textcolor{pink}{\bigstar})$, $n=4.20(\textcolor{yellow}{\bigstar})$, $n=6.60(\textcolor{cyan}{\bigstar})$, $n=5.00(\textcolor{green}{\bigstar})$, $n=5.40(\textcolor{brown}{\bigstar})$, $n=5.80(\textcolor{gray}{\bigstar})$,
$n=6.20(\textcolor{orange!50}{\bigstar})$, $n=6.60(\textcolor{magenta!50}{\bigstar})$, and $n=7.00(\textcolor{black}{\bigstar})$.}
\end{figure}
The balancing behavior of the forces $AF,\;HF,\;GF,\;\&~EF$ shows the stability of stellar objects. In this present case, the forces $AF,\;HF,\;GF,\;\&~EF$ are seen balanced to each other for two different sest of observational data, i.e., $(M=1.77, \;R_{b}=9.56)$ and $(M=1.97, \;R_{b}=10.3)$. The balancing development of the forces can be recognized from the Fig. (\textbf{14}). The balancing nature of $\mathscr{F}_{\mathrm{a}}$, $\mathscr{F}_{\mathrm{h}}$, $\mathscr{F}_{g}$, and $\mathscr{F}_{\mathrm{e}}$ suggests that our acquired results for two different stellar models are stable and physically acceptable.
\subsubsection{Stability analysis with causality condition}
Two famous velocities provide the causality analysis for the stability of stellar objects. Both of these velocities are known as radial and tangential speeds of sound, both are mentioned in this current study as $(SoS)_{r}=v_{r}\;\&\;(SoS)_{t}=v_{t}$. Both speeds of sounds are defined as:
\begin{eqnarray*}
v_{r}= \sqrt{\frac{dp_{r}}{dr} \times \frac{dr}{d\rho}}\;\;\;\;\;\;\Rightarrow  v^{2}_{r}= \frac{dp_{r}}{dr} \times \frac{dr}{d\rho}=\frac{dp_{r}}{d\rho},\\
v_{t}= \sqrt{\frac{dp_{t}}{dr} \times \frac{dr}{d\rho}}\;\;\;\;\;\;\Rightarrow  v^{2}_{t}= \frac{dp_{t}}{dr} \times \frac{dt}{d\rho}=\frac{dp_{t}}{d\rho}.
\end{eqnarray*}
Both are calculated  as
\begin{eqnarray}
v^{2}_{r}=\frac{dp_{r}}{d\rho}&&=-\frac{-1}{D_5^2 v_1 \left(Y r^2+1\right)^2 \sqrt{F_1(r)} F_5(r){}^2}\bigg(D_5 r \left(Y r^2+1\right) F_5(r) \left(B c F(r) \left(X r^2+1\right)^{n-1} \left(D_2 \left(X r^3+r\right)\right.\right.\nonumber\\&&\left.\left.+F_4(r) \left(X (2 n+1) r^2+1\right)\right)+Y n F_2(r){}^{n/2} \left(\left(D_8+D_9\right) \left(Y r^2+1\right)+2 Y r F_3(r)\right)-D_6+D_{10}\right)\nonumber\\&&-\frac{D_1 D_3 D_5 F_5(r)}{X r^2+1}-2 D_3 D_4 r^2 \left(Y r^2+1\right) F_5(r)-2 Y D_3 D_5 r^2 F_5(r)+D_3 D_5 \left(Y r^2+1\right) F_5(r)\nonumber\\&&-D_3 D_5 \left(D_{11}+D_{12}\right) r \left(Y r^2+1\right)\bigg),\label{49}\\
v^{2}_{t}=\frac{dp_{t}}{d\rho}&&=\frac{-1}{\left.v_1 \left(X r^2+1\right)^2 \left(Y r^2+1\right)^2 \sqrt{F_1(r)} F_5(r){}^2 F_6(r){}^3\right)}\bigg(F_6(r) \left(D_7 \left(F_5(r) \left(3 X Y r^4+r^2 (X+Y)+D_1-1\right)\right.\right.\nonumber\\&&\left.\left.+\left(D_{14}+D_{15}\right) r \left(X r^2+1\right) \left(Y r^2+1\right)\right)-r \left(Y r^2+1\right) F_5(r) \left(D_{16} \left(X r^2+1\right) F_2(r){}^{n/2} \left(-2 B c D_{21} r^2 \right.\right.\right.\nonumber\\&&\times\left.\left.\left.\left(X r^2+1\right)^n+A \left(D_{18}+D_{19}+D_{20}\right) \sqrt{F_1(r)}+D_{17}\right)+B c F(r) \left(X r^2+1\right)^n \left(D_{23} \left(X r^3+r\right)+F_7(r) \right.\right.\right.\nonumber\\&&\times\left.\left.\left.\left(X (2 n+1) r^2+1\right)\right)+D_{22} \left(X r^2+1\right)\right)\right)+4 D_4 D_7 r^2 \left(X r^2+1\right) \left(Y r^2+1\right) F_5(r)\bigg),\label{50}
\end{eqnarray}
where
\begin{eqnarray*}
v_1=\frac{\left(X r^2+1\right)^{n-2} \left(-2 c r F_0(r) F_6(r){}^2 \left(2 D_4 \left(X r^2+1\right)-X D_5 (n-1)\right)-2 c D_{13} D_5^3 r \left(X r^2+1\right)\right)}{D_5^3 F_6(r){}^2}+K Q.
\end{eqnarray*}
\begin{figure}
\centering \epsfig{file=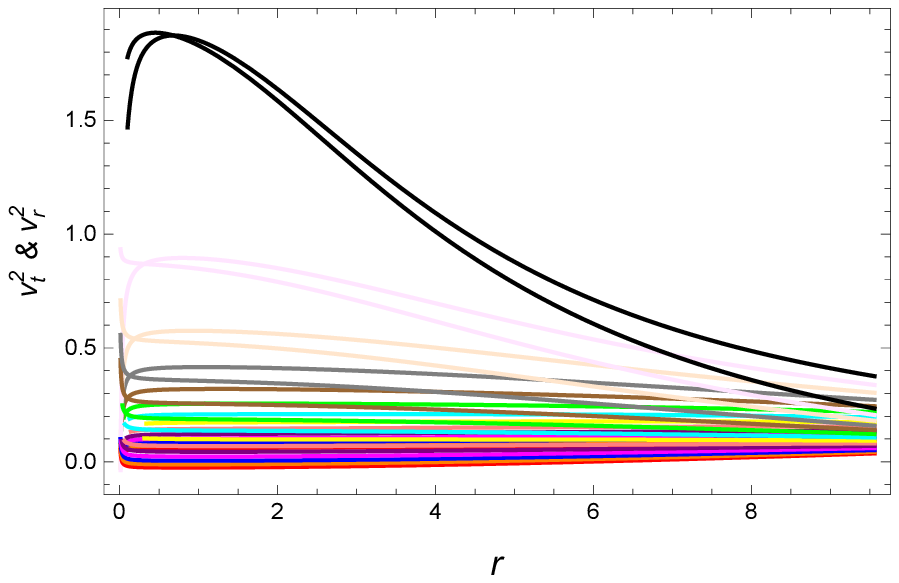, width=.48\linewidth,
height=2.1in}\epsfig{file=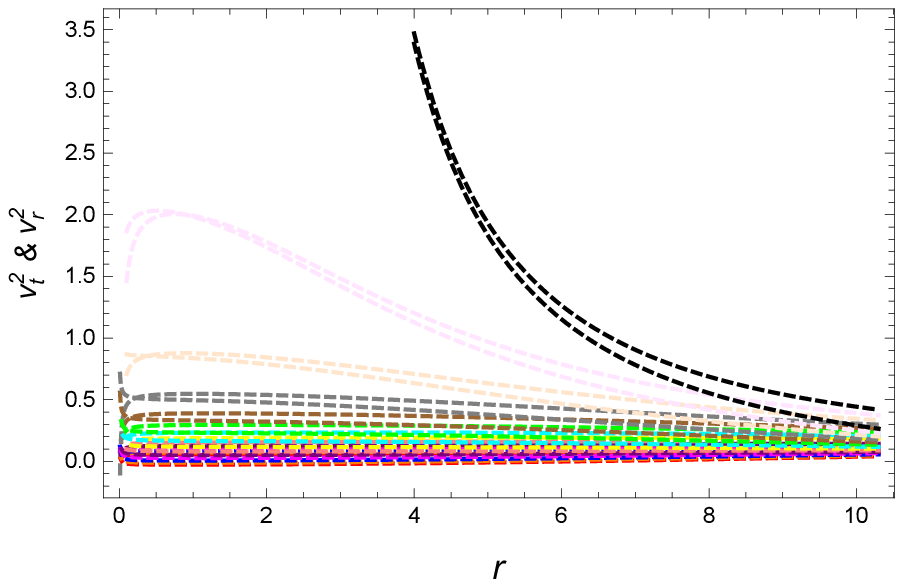, width=.48\linewidth,
height=2.1in} \caption{\label{Fig.15} describe the behavior of causality condition for two different models under $(M=1.77, \;R_{b}=9.56)$ and $(M=1.97, \;R_{b}=10.3)$ with $n=1.80(\textcolor{red}{\bigstar})$, $n=2.20(\textcolor{orange}{\bigstar})$, $n=2.60(\textcolor{blue}{\bigstar})$, $n=3.00(\textcolor{magenta}{\bigstar})$, $n=3.40(\textcolor{purple}{\bigstar})$, $n=3.80(\textcolor{pink}{\bigstar})$, $n=4.20(\textcolor{yellow}{\bigstar})$, $n=6.60(\textcolor{cyan}{\bigstar})$, $n=5.00(\textcolor{green}{\bigstar})$, $n=5.40(\textcolor{brown}{\bigstar})$, $n=5.80(\textcolor{gray}{\bigstar})$,
$n=6.20(\textcolor{orange!50}{\bigstar})$, $n=6.60(\textcolor{magenta!50}{\bigstar})$, and $n=7.00(\textcolor{black}{\bigstar})$.}
\end{figure}
\begin{figure}
\centering \epsfig{file=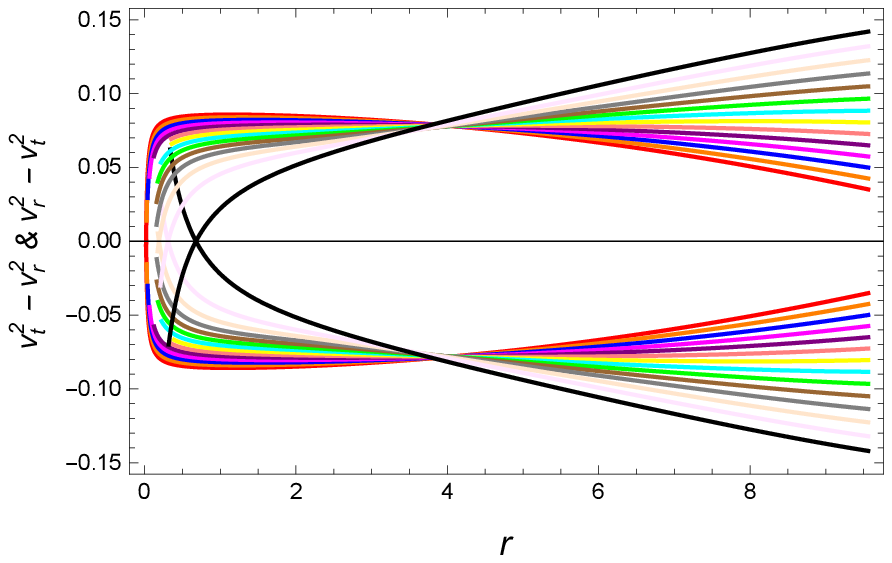, width=.48\linewidth,
height=2.1in}\epsfig{file=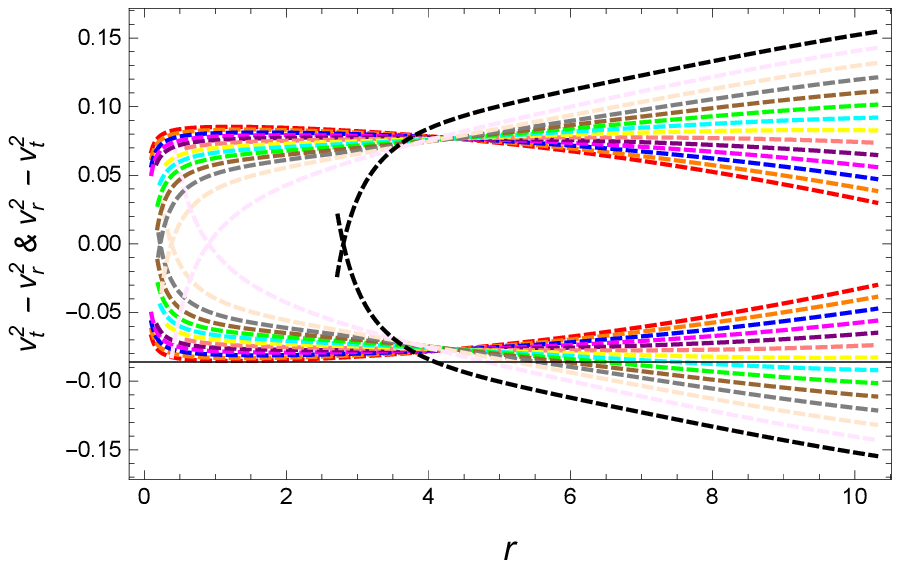, width=.48\linewidth,
height=2.1in} \caption{\label{Fig.16} describe the Abrea condition for two different models under $(M=1.77, \;R_{b}=9.56)$ and $(M=1.97, \;R_{b}=10.3)$ with $n=1.80(\textcolor{red}{\bigstar})$, $n=2.20(\textcolor{orange}{\bigstar})$, $n=2.60(\textcolor{blue}{\bigstar})$, $n=3.00(\textcolor{magenta}{\bigstar})$, $n=3.40(\textcolor{purple}{\bigstar})$, $n=3.80(\textcolor{pink}{\bigstar})$, $n=4.20(\textcolor{yellow}{\bigstar})$, $n=6.60(\textcolor{cyan}{\bigstar})$, $n=5.00(\textcolor{green}{\bigstar})$, $n=5.40(\textcolor{brown}{\bigstar})$, $n=5.80(\textcolor{gray}{\bigstar})$,
$n=6.20(\textcolor{orange!50}{\bigstar})$, $n=6.60(\textcolor{magenta!50}{\bigstar})$, and $n=7.00(\textcolor{black}{\bigstar})$.}
\end{figure}
The Fig. (\textbf{15}) illustrates the graphical representation of $(SoS)_{r}=v_{r}\;\&\;(SoS)_{t}=v_{t}$. Both the speeds of sounds satisfy the required condition, i.e., $0\leq(SoS)_{r}=v_{r}\;\&\;(SoS)_{t}=v_{t}<1$ for $n\in[1.8,\;6.2]-\{2,4,6\}$. Another important condition, i.e., $-1\leq v^{2}_{t}-v^{2}_{r}\leq 0$, which was presented by Abrea \cite{abre} is considered important. This condition $-1\leq v^{2}_{t}-v^{2}_{r}\leq 0$ is also satisfied for $n\in[1.8,\;6.2]-\{2,4,6\}$. The graphic analysis of Abrea condition can be perceived from the Fig. (\textbf{16}).

\begin{center}
\begin{table}
\caption{\label{tab1}{Summary of EoS parameters, equilibrium forces, energy conditions, and mass, compactness and red-shift functions.}}
\begin{tabular}{|c|c|c|c|c|c|c|c|c|}
    \hline
\multicolumn{4}{|c|}{$(M=1.77, \;R_{b}=9.56)\;\&\;(M=1.97, \;R_{b}=10.3)$}\\
    \hline
$n$        & $v^{2}_{r}\&v^{2}_{t}$                           &$v^{2}_{t}-v^{2}_{r}$                         & $\Gamma_{r}$\\
\hline
1.80     & $0\leq v^{2}_{r}\&v^{2}_{t}<1$                   &$-1\leq v^{2}_{t}-v^{2}_{r}\leq 0\;\& 0 \leq v^{2}_{r}-v^{2}_{t}\leq 1$            &$\Gamma_{r}>4/3$\\

2.20    & $0\leq v^{2}_{r}\&v^{2}_{t}<1$                   &$-1\leq v^{2}_{t}-v^{2}_{r}\leq 0\;\& 0 \leq v^{2}_{r}-v^{2}_{t}\leq 1$             &$\Gamma_{r}>4/3$\\

2.60    & $0\leq v^{2}_{r}\&v^{2}_{t}<1$                   &$-1\leq v^{2}_{t}-v^{2}_{r}\leq 0\;\& 0 \leq v^{2}_{r}-v^{2}_{t}\leq 1$            &$\Gamma_{r}>4/3$\\

3.00    & $0\leq v^{2}_{r}\&v^{2}_{t}<1$                   &$-1\leq v^{2}_{t}-v^{2}_{r}\leq 0\;\& 0 \leq v^{2}_{r}-v^{2}_{t}\leq 1$           &$\Gamma_{r}>4/3$\\

3.40   & $0\leq v^{2}_{r}\&v^{2}_{t}<1$                   &$-1\leq v^{2}_{t}-v^{2}_{r}\leq 0\;\& 0 \leq v^{2}_{r}-v^{2}_{t}\leq 1$           &$\Gamma_{r}>4/3$\\

3.80   & $0\leq v^{2}_{r}\&v^{2}_{t}<1$                   &$-1\leq v^{2}_{t}-v^{2}_{r}\leq 0\;\& 0 \leq v^{2}_{r}-v^{2}_{t}\leq 1$            &$\Gamma_{r}>4/3$\\

4.20    & $0\leq v^{2}_{r}\&v^{2}_{t}<1$                   &$-1\leq v^{2}_{t}-v^{2}_{r}\leq 0\;\& 0 \leq v^{2}_{r}-v^{2}_{t}\leq 1$          &$\Gamma_{r}>4/3$\\

4.60   & $0\leq v^{2}_{r}\&v^{2}_{t}<1$                   &$-1\leq v^{2}_{t}-v^{2}_{r}\leq 0\;\& 0 \leq v^{2}_{r}-v^{2}_{t}\leq 1$            &$\Gamma_{r}>4/3$\\

5.00  & $0\leq v^{2}_{r}\&v^{2}_{t}<1$                   &$-1\leq v^{2}_{t}-v^{2}_{r}\leq 0\;\& 0 \leq v^{2}_{r}-v^{2}_{t}\leq 1$           &$\Gamma_{r}>4/3$\\

5.40    & $0\leq v^{2}_{r}\&v^{2}_{t}<1$                   &$-1\leq v^{2}_{t}-v^{2}_{r}\leq 0\;\& 0 \leq v^{2}_{r}-v^{2}_{t}\leq 1$          &$\Gamma_{r}>4/3$\\

5.80    & $0\leq v^{2}_{r}\&v^{2}_{t}<1$                   &$-1\leq v^{2}_{t}-v^{2}_{r}\leq 0\;\& 0 \leq v^{2}_{r}-v^{2}_{t}\leq 1$            &$\Gamma_{r}>4/3$\\

6.20    & $0\leq v^{2}_{r}\&v^{2}_{t}<1$                   &$-1\leq v^{2}_{t}-v^{2}_{r}\leq 0\;\& 0 \leq v^{2}_{r}-v^{2}_{t}\leq 1$             &$\Gamma_{r}>4/3$\\

6.60   & Not satisfy the required condition                   &Not satisfy the required condition             &$\Gamma_{r}>4/3$\\

7.00    & Not satisfy the required condition                   &Not satisfy the required condition            &$\Gamma_{r}>4/3$\\
\hline
\end{tabular}
\end{table}
\end{center}
\subsection{Relativistic adiabatic index}
The product of two specific heats, argued by Steinmetz and Hillebrandt \cite{Hil}, provides the complete concept of relativistic adiabatic index, which is defined as
\begin{equation}\label{51}
\Gamma_{r}=\frac{p_r+\rho }{p_r}\times\frac{dp_r}{d\rho}.
\end{equation}
\begin{figure}
\centering \epsfig{file=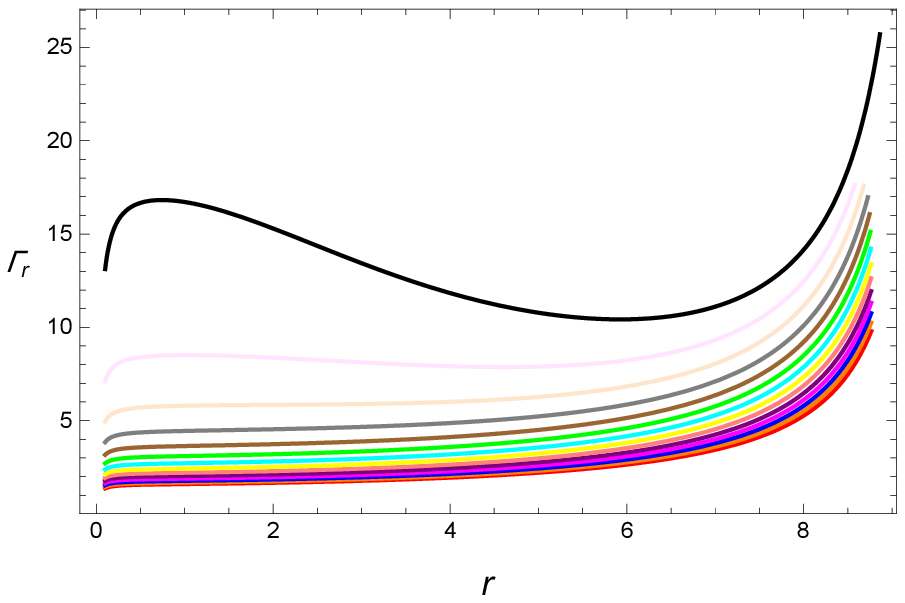, width=.48\linewidth,
height=2.1in}\epsfig{file=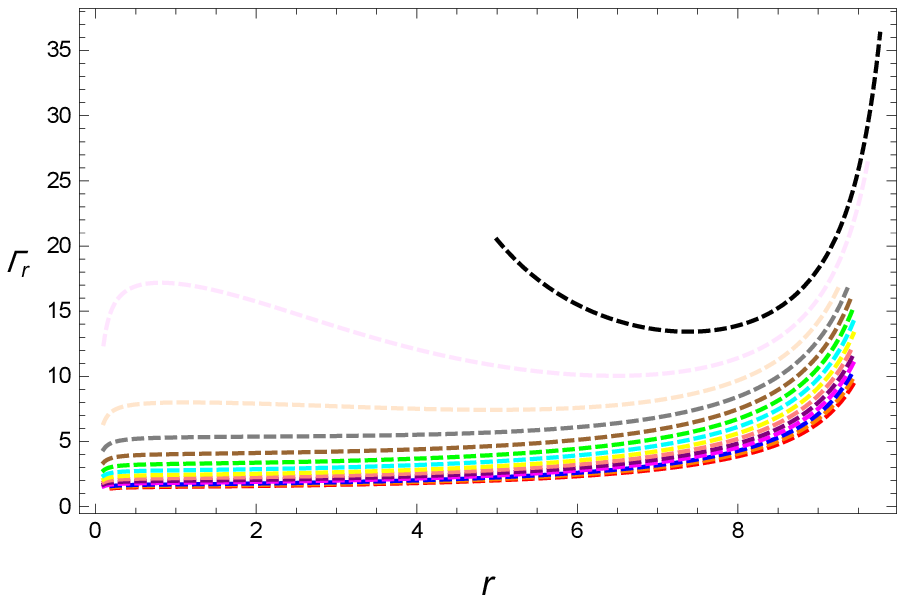, width=.48\linewidth,
height=2.1in} \caption{\label{Fig.17} shows the graphic development of adiabatic index for two different models under $(M=1.77, \;R_{b}=9.56)$ and $(M=1.97, \;R_{b}=10.3)$ with $n=1.80(\textcolor{red}{\bigstar})$, $n=2.20(\textcolor{orange}{\bigstar})$, $n=2.60(\textcolor{blue}{\bigstar})$, $n=3.00(\textcolor{magenta}{\bigstar})$, $n=3.40(\textcolor{purple}{\bigstar})$, $n=3.80(\textcolor{pink}{\bigstar})$, $n=4.20(\textcolor{yellow}{\bigstar})$, $n=6.60(\textcolor{cyan}{\bigstar})$, $n=5.00(\textcolor{green}{\bigstar})$, $n=5.40(\textcolor{brown}{\bigstar})$, $n=5.80(\textcolor{gray}{\bigstar})$,
$n=6.20(\textcolor{orange!50}{\bigstar})$, $n=6.60(\textcolor{magenta!50}{\bigstar})$, and $n=7.00(\textcolor{black}{\bigstar})$.}
\end{figure}
The above relativistic parameter $\Gamma_{r}$ is used to discuss the stability of the system.
Now, we are defined a condition for anisotropic matter profile, which is described as:
\begin{equation}\label{52}
\Gamma_{r}= \frac{4}{3}+\left(\frac{\kappa}{2}\frac{\rho_{c}p_{rc}}{|p^{'}_{rc}|}r+\frac{4}{3}\frac{(p_{tc}-p_{rc})}{|p^{'}_{rc}|r}\right)
\end{equation}
where $\kappa$ is an unknown and $p_{tc}, \;p_{rc}$ and $\rho_{c}$, are regarded as the initial tangential, radial components of pressure distribution and energy density function.
The parameter $\Gamma_{r}$ discusses the stability of Newtonian anisotropic sphere with condition $\Gamma_{r}>\frac{4}{3}$. An unstable sphere with anisotropic matter is observed when $\Gamma_{r}<\frac{4}{3}$.  In present study, we observe from the Fig. (\textbf{17}) that the parameter $\Gamma_{r}$ remains greater than $\frac{4}{3}$ throughout the configurations for $n\in[1.8,\;7]-\{2,4,6\}$. This condition shows that our acquired results are potentially stable.

\section{Concluding Remarks}

In this work, we have presented a new family of solutions of embedding
class-I model to discuss the physically viable solutions for two different
models of compact star in the presence of electric charge by employing the Bardeen
black hole geometry. In this context, we have considered the well-known Karmarkar
condition in the framework of Pandey-Sharma condition. In particular, we have
investigated some compulsory physical features related to the stellar configuration
graphically and analytically. For the current discussion, we use a spherically symmetric
spacetime with the charge matter involving anisotropic pressure sources. Here, we have
employed $e^{\lambda(r)}=\frac{c r^2 \left(X r^2+1\right)^n}{\left(Y r^2+1\right)^2}+1,$
specific model for $g_{rr}$ metric function. It is perceived that the parameter $n$ has
an important role in this current scenario. The specific values $n=2,~4,~6$ can not
be assumed for current study. It is also verified that some necessary physical
features are also not fulfilled for $n=7$. It is revealed that the calculated solutions
for both the compact stars models exhibit stable and realistic behavior for
$n\in[1.8,7)-\{2,4,6\}$. The graphical analysis of different physical features
have been presented in Figures \textbf{1}-\textbf{17}. In the following, we shall summarize our main results.
\begin{itemize}
\item From Fig. \textbf{1}, it can be checked that both the metric functions, i.e., $g_{tt}=e^{\nu}$, and $g_{rr}=e^{\lambda}$,
are calculated as $e^{\lambda(r =0)}=1$ and $e^{\nu(r =0)}\neq0$, which shows the physical supremacy of this current model.
Moreover, the Zeldovich's condition, i.e., $p_{rc}/ \rho_{c}=p_{tc}/ \rho_{c}\leq1$, is satisfied, which can be confirmed from Table-\textbf{III}.
\item The energy density function remains positive throughout the stellar interior for both of the compact stars models. The physical analysis of energy density function can be seen from Fig. \textbf{2}. Both the radial and tangential pressure components also remain positive throughout the stellar interior
for both models of compact star. The graphical behavior of both the pressure components can be seen in Figs \textbf{3-4}.
The nature of the charge profile and charge density can be noticed from the Figs. \textbf{5-6}.
\item It is found that the measure of anisotropic pressure $\triangle(r)$ is positive throughout the region of both the compact
stars and consequently, it supports the structure of compact star which has been provided in Fig. \textbf{7}.
The derivative of energy density function, radial pressure, and tangential pressure with respect to radial coordinate $r$ are
calculated as negative and their graphic nature can be confirmed form Fig. \textbf{8} and Table-\textbf{III}.
All the energy conditions, i.e., $\rho$, $p_r$, $p_t$, $\rho-p_{r}$, $\rho-p_{t}$, $\rho-p_{r}-2p_{t}$, are
observed to be satisfied for this model. Their graphical illustration has been provided in Figure \textbf{9}.
\item We have defined two EoS, namely $w_r$ and $w_t$ in radial and tangential directions respectively.
It is worthwhile to mention here that the constraints $w_i$ involved in the construction of $w_r$ and $w_t$ play an important role in exploring the nature of stellar structures. In our case, we can obtain $-1<\omega_r<-1/3$ for fixing the suitable values of $w_1$. Thus Bardeen stellar structures with Karmarkar Condition may support so-called dark stellar structures \cite{20Bhar5}-\cite{22Rahaman4}. Moreover, with the chosen constraints in the present study it is noticed that the values of these EoS parameters remain positive inside the stellar
object and also both values are calculated as less than 1, which can be seen from Fig. \textbf{10} and Table-\textbf{IV}.
\item The mass-radii function $m(r)$ remains positive, increasing and regular. Its graphical behavior can be observed from Table-\textbf{IV} and Fig. \textbf{11}. The compactness parameter $u(r)$ is remained positive and it also satisfies the Buchdahl limit, i.e., $u(r)\leq 8/9$ which can be seen
from Table-\textbf{IV} and Fig. \textbf{12}.
From Table-\textbf{IV} and Fig. \textbf{13}, it is observed that the surface red-shift function, i.e., $Z_{s}$ turns out to be zero at $r=0$
and gradually increases with the increase in radial coordinate. Also, it satisfies the Bohmer and Harko condition
under the anisotropic configuration, i.e., $Z_{s}<5$.
\item  From Table-\textbf{IV} and Fig. \textbf{14}, it is found that the forces $\mathscr{F}_{\mathrm{a}},
\;\mathscr{F}_{\mathrm{h}}$, $\mathscr{F}_{e}$, and $\mathscr{F}_{g}$ are consistent with the
equilibrium condition. It is also observed that these forces balance each other's effect and hence left the whole configuration stable.
It is expressed from Table-\textbf{V} and Figures \textbf{15-16} that the radial and tangential speeds of sound
for compact stars, which are denoted by $v^{2}_{r}$ and $v^{2}_{t}$ satisfy
the necessary condition for $1.8\leq n<6.6$, while for $n=6.6,~7$, these radial and tangential velocities are observed greater than 1,
and consequently, violates the necessary condition. Further, the causality stability condition
for these two values of $n$ is also violated for the present model. The more details can be seen from Table-\textbf{V}.
As far as adiabatic index is concerned, it is found from Table-II and Fig. \textbf{11} and Table-\textbf{V} that the adiabatic index
$\Gamma_{r}$ satisfies the inequality $\Gamma_{r}>4/3$ and has shown increasing behavior.
\end{itemize}
Hence, being sum-up, it can be concluded that our proposed models exhibit well-behaved nature and are physically considerable
for $n\in[1.8,7)-\{2,4,6\}$ as all the results coincide with \cite{jim,jim1}. For other choices of parameter $n$, the solutions do not favor a realistic compact star model.
%\\\\
%\textbf{Acknowledgements}\\\\
%Many thanks to the anonymous reviewers for valuable comments and suggestions to improve the paper. Mustafa acknowledges the support by the National Natural Science Foundation of China under Grants No. 11271008, No. 61072147, and No. 11975145. Shamir and Ahmad acknowledge National University of Computer and Emerging Sciences (NUCES) for research reward program.
\section*{Appendix (\textbf{I})}
\begin{eqnarray*}
F_0(r)&&=-\left(Y r^2+1\right) \left(r^2 \left(X \left(Y (2 n-1) r^2+2 n+3\right)-Y\right)+3\right)-c r^2 \left(X r^2+1\right)^{n+1},\\
F_1(r)&&=\frac{c r^2 \left(X r^2+1\right)^n}{\left(Y r^2+1\right)^2},\;\;\;\;\;F_2(r)=\frac{X Y r^2+Y}{X Y r^2+X},\\
F_3(r)&&=A \sqrt{F_1(r)} \left(c \left(X r^2+1\right)^n \left(K Q r^3-1\right)+K Q r \left(Y r^2+1\right)^2\right)+2 B c r \left(X r^2+1\right)^n,
\end{eqnarray*}
\begin{eqnarray*}
F_4(r)&&=c \left(X r^2+1\right)^n \left(K Q r^3-1\right)+K Q r \left(Y r^2+1\right)^2,~~
F_5(r)=A Y n r F_2(r){}^{n/2}+B \left(Y r^2+1\right) F(r) \sqrt{F_1(r)},\\
F_6(r)&&=c r^2 \left(X r^2+1\right)^n+Y^2 r^4+2 Y r^2+1,\\
F_7(r)&&=c \left(Y r^2+1\right) \left(X r^2+1\right)^n \left(X r^2 \left(Y n r^2-Y r^2+n+1\right)-Y r^2+1\right)+K Q r \left(X r^2+1\right) F_6(r){}^2,\\
F_8(r)&&=X r^2 \left(c r^2 \left(X r^2+1\right)^n+n \left(Y r^2+1\right)^2+2 Y r^2+2\right)+c r^2 \left(X r^2+1\right)^n+2 Y r^2+2,\\
F_9(r)&&=A \sqrt{F_1(r)} F_7(r)-B c r F_8(r) \left(X r^2+1\right)^n.
\end{eqnarray*}
\section*{Appendix (\textbf{II})}
\begin{eqnarray*}
D_1&&=X r^2 \left(Y n r^2-Y r^2+n+1\right)-Y r^2+1,\\
D_2&&=3 c K Q r^2 \left(X r^2+1\right)^n+2 X c n r \left(X r^2+1\right)^{n-1} \left(K Q r^3-1\right)+K Q \left(Y r^2+1\right)^2+4 Y K Q r^2 \left(Y r^2+1\right),\\
D_3&&=B c r F(r) F_4(r) \left(X r^2+1\right)^n+Y n \left(Y r^2+1\right) F_3(r) F_2(r){}^{n/2},\\
D_4&&=c \left(X (n+1) r^2+1\right) \left(X r^2+1\right)^{n-1}+2 Y^2 r^2+2 Y,\;\;D_5=c r^2 \left(X r^2+1\right)^n+\left(Y r^2+1\right)^2,\\
D_6&&=\frac{Y n^2 r (Y-X) F_2(r){}^{n/2} \left(A \sqrt{F_1(r)} \left(K Q r F_6(r)-c \left(X r^2+1\right)^n\right)+2 B c r \left(X r^2+1\right)^n\right)}{X r^2+1},\\
D_7&&=B c r F(r) F_7(r) \left(X r^2+1\right)^n+Y n \left(Y r^2+1\right) F_9(r) F_2(r){}^{n/2},\\
D_8&&=4 X B c n r^2 \left(X r^2+1\right)^{n-1}+2 B c \left(X r^2+1\right)^n+A D_2 \sqrt{F_1(r)},\\
D_9&&=\frac{A D_1 \sqrt{F_1(r)} \left(K Q r F_6(r)-c \left(X r^2+1\right)^n\right)}{r \left(X r^2+1\right) \left(Y r^2+1\right)},\;\;D_{10}=\frac{X Y B c n r^2 F_4(r) \left(X r^2+1\right)^n \left(F_2(r){}^{n/2}-F(r)\right)}{X Y r^2+X},\\
D_{11}&&=-\frac{A Y n^2 r^2 (Y-X) F_2(r){}^{n/2}}{\left(X r^2+1\right) \left(Y r^2+1\right)}+A Y n F_2(r){}^{n/2}+Y B n r \sqrt{F_1(r)} \left(F_2(r){}^{n/2}-F(r)\right),\\
D_{12}&&=\frac{B D_1 F(r) \sqrt{F_1(r)}}{X r^3+r}+2 Y B r F(r) \sqrt{F_1(r)},\\
D_{13}&&=X \left(n \left(c r^2 \left(X r^2+1\right)^n+6 Y^2 r^4+8 Y r^2+2\right)+2 c r^2 \left(X r^2+1\right)^n-3 Y^2 r^4+4 Y r^2+3\right)+c \left(X r^2+1\right)^n\\&&-2 Y^2 r^2+2 Y,\\
D_{14}&&=-\frac{A Y n^2 r^2 (Y-X) F_2(r){}^{n/2}}{\left(X r^2+1\right) \left(Y r^2+1\right)}+A Y n F_2(r){}^{n/2}+Y B n r \sqrt{F_1(r)} \left(F_2(r){}^{n/2}-F(r)\right),\\
D_{15}&&=\frac{B D_1 F(r) \sqrt{F_1(r)}}{X r^3+r}+2 Y B r F(r) \sqrt{F_1(r)},\\
D_{16}&&=\frac{Y^2 n^2 r (X-Y) F_9(r) F_2(r){}^{n/2}}{X Y r^2+Y}+2 Y^2 n r F_9(r) F_2(r){}^{n/2}+Y n \left(Y r^2+1\right),\\
D_{17}&&=\frac{A D_1 \sqrt{F_1(r)} F_7(r)}{r \left(X r^2+1\right) \left(Y r^2+1\right)}-2 X B c n r^2 F_8(r) \left(X r^2+1\right)^{n-1}-B c F_8(r) \left(X r^2+1\right)^n,\\
D_{18}&&=2 X c D_1 n r \left(Y r^2+1\right) \left(X r^2+1\right)^{n-1}+2 Y c D_1 r \left(X r^2+1\right)^n+K Q \left(X r^2+1\right) F_6(r){}^2,\\
D_{19}&&=2 c r \left(Y r^2+1\right) \left(X r^2+1\right)^n \left(X \left(2 Y n r^2-2 Y r^2+n+1\right)-Y\right),\\
D_{20}&&=2 X K Q \left(r^3 \left(c \left(X r^2+1\right)^n+2 Y\right)+Y^2 r^5+r\right)^2+4 D_4 K Q r^2 \left(X r^2+1\right) F_6(r),\\
D_{21}&&=X \left(n \left(c r^2 \left(X r^2+1\right)^n+3 Y^2 r^4+4 Y r^2+1\right)+2 c r^2 \left(X r^2+1\right)^n+4 Y r^2+2\right)+c \left(X r^2+1\right)^n+2 Y,\\
D_{22}&&=\frac{X Y B c n r^2 F_7(r) \left(X r^2+1\right)^n \left(F_2(r){}^{n/2}-F(r)\right)}{X Y r^2+X},
\end{eqnarray*}
\begin{eqnarray*}
D_{23}&&=2 X K Q \left(r^3 \left(c \left(X r^2+1\right)^n+2 Y\right)+Y^2 r^5+r\right)^2+2 X c D_1 n r \left(Y r^2+1\right) \left(X r^2+1\right)^{n-1}\\&&+2 Y c D_1 r \left(X r^2+1\right)^n+2 c r \left(Y r^2+1\right) \left(X r^2+1\right)^n \left(X \left(2 Y n r^2-2 Y r^2+n+1\right)-Y\right)\\&&+4 D_4 K Q r^2 \left(X r^2+1\right) F_6(r)+K Q \left(X r^2+1\right) F_6(r){}^2
\end{eqnarray*}
\section*{Appendix (\textbf{III})}
\begin{eqnarray*}
w_1&&=\frac{c \left(X r^2+1\right)^{n-1} \left(-\left(Y r^2+1\right) \left(r^2 \left(X \left(Y (2 n-1) r^2+2 n+3\right)-Y\right)+3\right)-c r^2 \left(X r^2+1\right)^{n+1}\right)}{\left(c r^2 \left(X r^2+1\right)^n+\left(Y r^2+1\right)^2\right)^2}+K Q r,\\
w_2&&=A \sqrt{F_1(r)} \left(K Q r \left(X r^2+1\right) \left(c r^2 \left(X r^2+1\right)^n+\left(Y r^2+1\right)^2\right)^2+c \left(Y r^2+1\right) \left(X r^2+1\right)^n\right.\\&&\times \left. \left(X Y (n-1)r^4+r^2 (X n+X-Y)+1\right)\right)+B c r \left(X r^2+1\right)^n \left(-\left(Y r^2+1\right) \left(X r^2 \left(Y n r^2+n+2\right)\right.\right.\\&&+ \left.\left.2\right)-c r^2 \left(X r^2+1\right)^{n+1}\right),\\
w_3&&=K Q r \left(X r^2+1\right) \left(c r^2 \left(X r^2+1\right)^n+\left(Y r^2+1\right)^2\right)^2+c \left(Y r^2+1\right) \left(X r^2+1\right)^n \left(X Y (n-1) r^4\right.\\&&+ \left.+r^2 (X n+X-Y)+1\right).
\end{eqnarray*}


\section*{References}
\begin{thebibliography}{36}

\bibitem{OV} J. R. Oppenheimer and G. Volkoff, Phys. Rev. \textbf{55}(1939) 374.
\bibitem{1Schwarz1} K. Schwarzschild, Sitz. Deut. Akad. Wiss. Berlin, Kl. Math. Phys. \textbf{189} (1916a).
\bibitem{2Schwarz2} K. Schwarzschild, Sitz. Deut. Akad. Wiss. Berlin, Kl. Math. Phys.  \textbf{424} (1916b).
\bibitem{3Bowers} R. L. Bowers and E. P. T. Liang, Astrophys. J. \textbf{188} (1974) 657-665.

\bibitem{4Bekenstei} J. D. Bekenstein, Phys. Rev. \textbf{D4} (1971) 2185-2190.

\bibitem{8Vaidya} P. C. Vaidya, Proc. Ind. Acad. Sci.  \textbf{A33} (1951) 264-276.
\bibitem {Santos} N. O. Santos, Mon. Not. R. Astron. Soc. \textbf{216} (1985) 403-410.

\bibitem{10Bonnor} W. B. Bonnor, A. K. G. de Oliveira and N.O. Santos, Phys. Rep. \textbf{181} (1989) 269—326.
\bibitem{11Sharma} R. Sharma and R. Tikekar, Pramana J. of Phys. \textbf{79} (2012) 501-509.

\bibitem{12Govende} M. Govender, et al, Astrophys. Space Sci. \textbf{361} (2016) 33.
\bibitem{13Bhar2} P. Bhar, Astrophys. Space Sci. \textbf{356} (2015) 309-318.
\bibitem{14Bhar3} P. Bhar, Eur. Phys. J. C: \textbf{75} (2015) 123.
\bibitem{15Singh4} K. N. Singh and N. Pant, Astrophys. Space Sci. \textbf{358} (2015) 44.
\bibitem{16Andreasson} H. Andreasson, J. Phys. Conference Series: \textbf{189} (2009) 012001.
\bibitem{17Takisa} T. P. Mafa, S. Ray and S. D. Maharaj, Astrophys. Space Sci. \textbf{350} (2014) 733.
\bibitem{18Ngubelanga} S. A. Ngubelanga, S. D. Maharaj and S. Ray, Astrophys. Space Sci. \textbf{357} (2015) 74.
\bibitem{19Govender1} M. Govender and S. Thirukkanesh, Astrophys. Space Sci,\textbf{358} (2015) 16-22.
\bibitem{20Bhar5} P. Bhar and F. Rahaman, Euro. Phys. J. \textbf{C75} (2015) 41.
\bibitem{21Lobofsn} F. S. N. Lobo, Class. Quantum Grav. \textbf{23} (2006) 1525.
\bibitem{22Rahaman4} F. Rahaman et al, Euro. Phys. J. \textbf{C72} (2012) 2071.
\bibitem{23Rahaman5} F. Rahaman et al, Gen. Relativ. Gravit. \textbf{44} (2012) 107-124.
\bibitem{24Bhar7} P. Bhar, Astrophys. Space Sci. \textbf{359} (2015) 41.
\bibitem{25Chavanis} P. H. Chavanis and T. Harko, Phys. Rev. \textbf{D86} (2012) 064011.
\bibitem{26Harko7} T. Harko, Phys. Rev. \textbf{D68} (2003) 064005.
\bibitem{27Dadhich} N. Dadhich et at, Phys. Rev. \textbf{D88} (2013) 084024.
\bibitem{34Karmarkar3} K. R. Karmarkar, Proc. Indian. Acad. Sci. \textbf{A27} (1948) 56-60.

\bibitem{E1} L. Schlai, Ann. di Mat. \textbf{5} (1871) 170.
\bibitem{E2} J. Nash, Ann. Math., \textbf{63}, (1956) 20.
\bibitem{E3} Y. K. Gupta and R. S. Gupta, Gen. Rel. Gray. \textbf{6}, (1986) 641.
\bibitem{E4} Y. K. Gupta and J. R. Sharma, Gen Relat Gravit \textbf{28} (1996) 1447.

\bibitem{KC} M. Kohler and K.L. Chao, Z. Naturforsch. Ser. \textbf{A20} (1965) 1537.

\bibitem{M1} S. K. Maurya, Y. K. Gupta, Saibal Ray and S. R. Chowdhury, Eur. Phys. J. \textbf{C75} (2015) 389.

\bibitem{M2} S. K. Maurya, Y. K. Gupta, T. T. Smitha and F. Rahaman, Eur. Phys. J. \textbf{A52} (2016) 191.

\bibitem{M3} P. Bhar, S. K. Maurya, Y. K. Gupta and T. Manna, Eur. Phys. J. \textbf{A52} (2016) 312.

\bibitem{M004} D. Deb, S. V. Ketov, S. K. Maurya MNRAS 485 (2019) 5652.

\bibitem{M00004} S. K. Maurya et al., Phys. Rev. D 100 (2019).

\bibitem{M4} S. K. Maurya, Y. K. Gupta, F. Rahaman, M. Rahaman and A. Banerjee, Annals of Physics \textbf{385} (2017) 532.

\bibitem{M5} S. K. Maurya, Y. K. Gupta, B. Dayanandan and S. Ray, Eur. Phys. J. \textbf{C76} (2016) 266.

\bibitem{M6} S. K. Maurya, Y. K. Gupta, S. Ray and D. Deb, Eur. Phys. J. \textbf{C77} (2016) 1.

\bibitem{28Joshi4} P. S. Joshi, and D. Malafarina, Int. J. Mod. Phys. \textbf{D20} (2011) 2641- 2729.
\bibitem{29Dadhich2} N. Dadhich et at,  Phys.Lett. \textbf{B711} (2012) 196-198.
\bibitem{30Hansraj1} S. Hansraj, et at, Eur. Phys. J. \textbf{C75} (2015) 277.
\bibitem{31Maharaj3} S. D. Maharaj et at, Phys. Rev. \textbf{D91} (2015) 084049.
\bibitem{32Dadhich3} N. Dadhich et at, Phys. Rev. \textbf{D93} (2016) 044072.
\bibitem{33Banerjee2} A. Banerjee et at, Euro. Phys. J. \textbf{C76}  (2016) 34.


\bibitem{Buchdahl} H. A. Buchdahl , Phys. Rev. \textbf{116} (1959) 1027.
\bibitem{Baade} W. Baade and F. Zwicky,  Phys. Rev. \textbf{46} (1934) 76.
\bibitem{Longair} M. S. Longair, \textit{High Energy Astrophysics} (Cambridge Univeristy Press, 1994).
\bibitem{Ghosh} P. Ghosh, \textit{Rotation and Accretion Powered Pulsars}(World Scientific, 2007).
\bibitem{Ruderman} R. Ruderman, Annu. Rev. Astron. Astrophys. \textbf{10} (1972) 427.
\bibitem{Maurya} S. K. Maurya  and Y.K. Gupta, Astrophys. Space Sci. \textbf{344} (2013) 243.
\bibitem{Maurya2} S. K. Maurya  and Y.K. Gupta, Phys. Scr. \textbf{86} (2012) 025009.

\bibitem{Maharaj} S. D. Maharaj, J. M. Sunzu and S. Ray, Eur. Phys. J. Plus \textbf{129} (2014) 3.
\bibitem{Kalam} M. Kalam  et al., Eur. Phys. J. \textbf{C72} (2012) 2248.
\bibitem{K&B} K.D. Krori  and J. Barua, J. Phys. A: Math. Gen. \textbf{8} (1975) 508.
\bibitem{Rahaman1} F. Rahaman et al., Gen. Relativ. Grav. \textbf{44} (2012) 107.
\bibitem{Rahaman2} F. Rahaman et al., Eur. Phys. J. C \textbf{72} (2012) 2071.
\bibitem{MaK} M. K. Mak and T. Harko, Int. J. Mod. Phys. \textbf{D13} (2004) 149.

\bibitem{Bekenstein} J. D. Bekenstein, Phys. Rev. \textbf{D4} (1971) 2185.

\bibitem{Zhang} J. L. Zhang, W. Y. Chau and T. Y. Deng, Astrophys. Space Sci. \textbf{88} (1982) 81.

\bibitem{Felice} F. de Felice, Y. Yu and Z. Fang, Mon. Not. R. Astron. Soc. \textbf{277} (1995) L17.

\bibitem{Felice1} F. de Felice, S. M. Liu and Y. Q. Yu, Class. Quantum Grav. \textbf{16} (1999) 2669.


\bibitem{Ref1} S.K. Maurya, Eur. Phys. J. C \textbf{79} (2019) 958.

\bibitem{Ref2} S.K. Maurya and T. Francisco, Phys. Dark Uni. 27 (2020) 100442.

\bibitem{Ref3} S.K. Maurya, A. Banerjee and P. Channuie, Chin. Phys. C 42 (2018) 055101.

\bibitem{Kumar} J. Kumar, S.K. Maurya, A.K. Prasad and A. Banerjee, JCAP \textbf{2019} (2019) 005.
\bibitem{Takisa} T. P. Mafa, S.D. Maharaj and L. L. Leeuw: Eur. Phys. J. \textbf{C79}(2019) 8.

\bibitem{2600} S. Rosseland, Mon. Not. R. Astron. Soc., \textbf{84} (1924) 720.

\bibitem{8143} R. Kippenhahm and A. Weigert, Stellar Structure and Evolution (Springer, Berlin, 1990).
\bibitem{9143} A. I. Sokolov, Zh. Eksp. Teor. Fiz. 79 (1980) 1137.





\bibitem{41} L.P. Eisenhart, \emph{Riemannian Geometry} (Princeton University Press, Princeton, (1966).

\bibitem{1000} K. Lake, Phys. Rev. D \textbf{67}, 104015 (2003).

\bibitem{Bardeen} J. M. Bardeen, Proceedings of GR-5, Tiflis, Georgia, U.S.S.R. page \textbf{174} (1968).
\bibitem{Garcia} E. Ayon-Beato, A. Garcia, Phys. Lett. B: \textbf{493} (2000) 149.

\bibitem{Moreno} C. Moreno and O. Sarbach, Phys. Rev. \textbf{D67} (2003) 024028.

\bibitem{Hawking} S. W. Hawking and G.F.R. Ellis, \textit{The large scale structure of space-time}  (Cambridge University Press, 1973)

\bibitem{Fernando1} S. Fernando, J. Correa, Phys. Rev. \textbf{D86}, 64039 (2012)
\bibitem{Fernando2} S. Fernando, Int. J. Mod. Phys. \textbf{D26}(2017)1750071.
\bibitem{Flachi} A. Flachi, J. Lemos, Phys. Rev. \textbf{D87}, 024034 (2013)
\bibitem{Ulhoa} S. C. Ulhoa: Braz, Jour. Phys. \textbf{44}, 380 (2014).
\bibitem{Nordstrom} G. Nordström, Verhandl. Koninkl. Ned. Akad. Wetenschap., Afdel. Natuurk., 26 (1918) 1201.
\bibitem{40} C.G. Bohmer and T. Harko, Class. Quantum Grav. \textbf{23}, 6479 (2006).
\bibitem{41} B. V. Ivanov, Phys. Rev. \textbf{D65}, 104001 (2002).
\bibitem{abre} H. Abreu, H. Hernandez, and L.A. Nunez, Class. Quantum Grav. \textbf{24}, 4631 (2007).
\bibitem{Hil} W. Hillebrandt and K. O. Steinmetz, Astron. Astrophys. \textbf{53}, 283 (1976).
\bibitem{jim} G. Mustafa et al, Annals of Physics \textbf{413} (2020) 168059.
\bibitem{jim1} G. Mustafa et al, Eur. Phys. J. C \textbf{80} (2020) 26.
\end{thebibliography}
\end{document}